\documentclass[12pt]{article}

\usepackage[ascii]{inputenc}
\usepackage{amsmath,amssymb,amsfonts,amsthm}
\usepackage[margin=2.8cm]{geometry}
\usepackage{graphicx}
\usepackage[caption=false]{subfig}
\usepackage[pdftex,bookmarks=false,colorlinks=true,linkcolor=blue,
citecolor=blue,filecolor=black,urlcolor=blue]{hyperref}

\usepackage{lmodern,bm}
\usepackage{latexsym}
\usepackage{longtable}
\usepackage{epsfig}
\usepackage{psfrag}
\numberwithin{equation}{section}

\begin{document}
\begin{center}
\vspace*{2cm} {\Large {\bf The Kardar-Parisi-Zhang  Equation - a Statistical Physics Perspective\bigskip\bigskip\\}}
{\large Herbert Spohn}\bigskip\bigskip\\
Zentrum Mathematik and Physik Department,
Technische Universit\"at M\"unchen,
Boltzmannstra{\ss}e 3, 85747 Garching, Germany. 
{\tt spohn@ma.tum.de}
\vspace{5cm}\end{center} 
 \newpage\noindent\\
Table of contents\\\\
1. Stable-metastable interface dynamics \\\\
2. Scaling properties, KPZ equation as  weak drive limit\\\\
3. Eden type growth models\\\\
4. The KPZ universality class\\\\
5. Directed polymers in a random medium\\\\
6. Replica solutions\\\\
7. Statistical mechanics of line ensembles\\\\
8. Noisy local conservation laws\\\\
9. One-dimensional fluids and anharmonic chains\\\\
10. Discrete nonlinear Schr\"{o}dinger equation\\\\
11. Linearized Euler equations\\\\
12. Linear fluctuating hydrodynamics\\\\
13. Second order expansion, nonlinear fluctuating hydrodynamics\\\\
14. Coupling coefficients, dynamical phase diagram\\\\
15. Low temperature DNLS\\\\
16. Mode-coupling theory\\\\
17. Molecular dynamics and other missed topics
\newpage\noindent
\section{Stable-metastable interface dynamics}
\label{sec1}
The ground breaking 1986 contribution of Kardar, Parisi, and Zhang, for short KPZ,  is entitled ``Dynamic scaling of growing interfaces'' \cite{KPZ86}. They study the dynamics of  an interface arising from a stable bulk phase in contact with a metastable one. Assuming that the two bulk phases have no conservation laws and relax exponentially fast,  KPZ argue that the motion of the interface is governed by the stochastic PDE
\begin{equation}\label{1.1}
\partial_t h = \tfrac{1}{2}\lambda (\nabla_{\!\boldsymbol{x}} h)^2  + \nu \Delta_{\boldsymbol{x}} h +\sqrt{D} \xi\,,
\end{equation}
where  $h(\boldsymbol{x},t)$ denotes the height function over the substrate space $\boldsymbol{x} \in \mathbb{R}^d$ at  time $t \geq 0$. The nonlinearity arises from the asymmetry between the two phases. At the interface a transition from metastable to stable is fast while the reverse process is strongly suppressed. The Laplacian reflects the interface tension and the space-time white noise,  $\xi(\boldsymbol{x},t)$,  models the randomness in transitions from metastable to stable. $\lambda, \nu, D$ are material parameters, following the original KPZ notation, $ \nu>0$, $D > 0$. 

Over the past fifteen years we have witnessed spectacular advances in the case of a two-dimensional bulk, one-dimensional interface, both from the experimental and theoretical side, see the reviews \cite{Jo05,Sp06, SaSp11a, FeSp11,QRev,CorwinRev,BoGo12,BoPe14,Ta15,QuSp15}. Therefore we restrict our discussions immediately to $d=1$, in which case Eq. \eqref{1.1} reads
\begin{equation}\label{1.1a}
\partial_t h = \tfrac{1}{2}\lambda (\partial_x h)^2  +\nu (\partial_x)^2 h +\sqrt{D} \xi\,.
\end{equation}
 In a certain sense as a trade-off, also much progress has been achieved on the less understood case of $2+1$  dimensions. The 
 long standing open theoretical problem of an upper critical dimension is in perspective again. 
 I refer to the recent contribution by T. Halpin-Healy and K. Takeuchi \cite{HaKa15}, which serves as a perfect  trail head and provides instructive details along the path. 

To have a concrete physical picture and a better understanding of the approximations underlying 
\eqref{1.1a}, it is illuminating to first consider the two-dimensional ferromagnetic Ising model  with Glauber spin flip dynamics, as one of the most basic model system of statistical mechanics. Its spin configurations are denoted by $\sigma = \{\sigma_{\boldsymbol{j}}, \boldsymbol{j} \in \mathbb{Z}^2\}$ with $\sigma_{\boldsymbol{j}} = \pm 1$.  The Ising energy is 
\begin{equation}\label{1.2}
H(\sigma) = - \sum_{\boldsymbol{i},\boldsymbol{j} \in \mathbb{Z}^2, |\boldsymbol{i}-\boldsymbol{j}|=1} \sigma_
{\boldsymbol{i}}\sigma_{\boldsymbol{j}} - h \sum_{\boldsymbol{j}\in \mathbb{Z}^2}\sigma_{\boldsymbol{j}}\,,\qquad
H_0(\sigma) = H(\sigma)|_{h=0}\,,
\end{equation}
where the first sum is over nearest neighbor pairs and the spin coupling is used as energy scale. The flip rate from $\sigma_{\boldsymbol{j}}$ to  $-\sigma_{\boldsymbol{j}}$ is given by 
\[
c_{\boldsymbol{j}}(\sigma) = \left\{\begin{array}{ll}
1, &\mbox{if $\Delta_{\boldsymbol{j}} H(\sigma) \leq 0$\,,}\smallskip\\
\mathrm{e}^{-\beta \Delta_{\boldsymbol{j}} H (\sigma)},& \mbox{if $\Delta_{\boldsymbol{j}} H(\sigma) > 0$\,.}
\end{array}
\right. \]
Here $\beta > 0$ is the inverse temperature and  $\Delta H_{\boldsymbol{j}}$ the energy difference in a spin flip at ${\boldsymbol{j}}$,     $\Delta H_{\boldsymbol{j}}(\sigma) = H(\sigma^{\boldsymbol{j}}) - H(\sigma)$, where $\sigma^{\boldsymbol{j}}$
equals $\sigma$ with $\sigma_{\boldsymbol{j}}$ flipped to $- \sigma_{\boldsymbol{j}}$.

First note that the bulk dynamics has no conservation law and, away from criticality, an exponentially fast relaxation. If there would be bulk conservation laws, the interface dynamics to be studied would have very different properties.

Considering  $H_0$ and  $\beta > \beta_\mathrm{c}$, $1/\beta_\mathrm{c}$ the critical temperature, the Glauber dynamics has exactly two (extremal)
invariant measures, denoted by $\mu_{\pm}$. The $\mu_{+}$ phase is obtained through the infinite volume limit of 
$Z^{-1}\mathrm{e}^{-\beta H_0}$ with $+$ boundary conditions and $\mu_{-}$ equals $\mu_{+}$ after a global spin flip.
$\mu_{+}$ has a strictly positive spontaneous magnetization. Both phases are stable and have exactly the same free energy. One could start however from a non-stationary initial state. A much studied example is a low temperature quench, for which the initial state  is equilibrium at $\beta = 0$, while the dynamics runs at $\beta \gg \beta_\mathrm{c}$. In our context we fix $\beta > \beta_\mathrm{c}$  and consider a set-up with two rather large, possibly macroscopic, disjoint domains $\Lambda_{+(-)}$ with smooth boundaries such that
$\Lambda_+ \cup\Lambda_{-} = \mathbb{Z}^2$. In $\Lambda_{+}$ we choose the state $\mu_{+}$ and
in $\Lambda_{-}$ the state $\mu_{-}$,
adopting some physically reasonable choice at the interface $\partial \Lambda_+ \cup \partial\Lambda_{-}$.
A standard example would be the half-spaces $\Lambda_{+(-)} = \{x, \vec{n}\cdot x \geq (<) \,0\}$ specified by the normal $\vec{n}$. Thereby a stable-stable interface is imposed. The interface is initially flat and remains sharply localized in the course of time under the Glauber dynamics for $h=0$, \textit{i.e.}  with flip rates derived from $H_0$, but develops fluctuations  with an amplitude of size $t^{1/4}$. Away from the interface the bulk has a statistics which in good approximation is described by either $\mu_{+}$ or $\mu_{-}$.

KPZ raised the issue of how such interface motion is modified when the Glauber dynamics is run at a small $h >0$. Then $\mu_+$ remains  stable, but $\mu_{-}$ has turned metastable. At the interface the Glauber dynamics easily flips a $-$ spin to a $+$ spin, while the reversed process is suppressed.
 Thereby the $+$ phase expands into the $-$ phase and the stable-metastable interface acquires a non-zero drift velocity. 
 In addition, inside the metastable $-$ domain a stable $+$ nucleus could be formed, either far out statically or through a dynamical fluctuation. Such an event is unlikely, but once it happens the stable nucleus will grow and possibly collide with the already present interface. Our description will be restricted to times before such a collision. In fact,  in most models such extra nucleation events are suppressed entirely. In the course of time
 the stable-metastable interface remains well
 localized, but roughens on top of the systematic motion. Our goal is to understand the space-time statistical properties of this roughening process. Note that the effective interface dynamics is no longer invariant under time-reversal, in contrast to the underlying Glauber dynamics. As an additional issue of great interest, we have naturally arrived at a stochastic field theory
 with a non-symmetric generator.   

A theoretical study of the Glauber dynamics with such initial conditions seems to be exceedingly difficult. Fortunately,
in the limit of zero temperature one arrives at tractable models. As a general consensus, one expects that the large scale
properties of the interface will not change when heating up, of course always staying below the critical temperature. 

More specifically let us start from an interface given through a down-right lattice path. Below that path all spins are up and above they are down. 
At zero temperature only flips with $\Delta_{\boldsymbol{j}} H \leq 0$ are admissible. Under this constraint a $-$ spin flips to $+$ with rate 
$p$ while the reversed transition occurs  with rate $q$, $p+q =1$ to fix the time scale. $h=0$ corresponds to  $p=q =\tfrac{1}{2}$. The dynamics is stochastically reversible. On the other hand, for small $h >0$ the stable-metastable flip rates differ, $p > q$, and the stochastic dynamics  is non-reversible.
 The interface width is one lattice unit.
The zero temperature Glauber dynamics never leaves the set of down-right paths. Thus, in the spirit of the KPZ equation, we have accomplished an autonomous interface dynamics.

The conventional scheme to define a height function, $h(j,t)$ with $j \in \mathbb{Z}$, is to choose  the anti-diagonal as reference line.   The height function satisfies the constraint $|h(j+1,t) - h(j,t)| = 1$. We draw $h(j,t)$ as a continuous broken line with slope $\pm 1$, $\diagup$, $\diagdown$,
such that $h(j,t)$ is at the lattice points $\mathbb{Z} + \tfrac{1}{2} $. Under the height dynamics, independently
a local minimum of $h$, $\diagdown\diagup$, flips  to $\diagup\diagdown$  with rate $p$ and thus $h(j) \Rightarrow h(j) +2$. Correspondingly a local maximum, $\diagup\diagdown$, flips to $\diagdown\diagup$ with rate $q$ and thus $h(j) \Rightarrow h(j) - 2$. Since the slope takes only values $\pm1$, this height dynamics goes under the label ``single-step". $p=q$ is the symmetric dynamics, corresponding to a stable-stable interface, and $p\neq q$ is the asymmetric case, including the totally asymmetric limits $p= 1$, $q=1$. 

For an interface parallel to one of the lattice axes, according to our rules the interface cannot move, no flip is allowed. 
To arrive at a non-trivial dynamics the limit $\beta \to \infty$ has to be taken differently. As initial configuration let us assume that all spins in the upper half plane are down and are up in the lower half plane. Then at low temperatures the slow process are flips from $-$ to $+$
with $\Delta_{\boldsymbol{j}} H = 2$. Once this has happened the allowed flips with $\Delta_{\boldsymbol{j}} H = 0$ are fast. Under a suitable scaling one arrives at the polynuclear growth (PNG) model. The height function is $h(x,t)$ with space $x\in \mathbb{R}$ and time $t \geq 0$. $ x \mapsto h(x,t)$ is piecewise constant and takes values in $\mathbb{Z}$ such that  up-steps are of size $1$ and down-steps  of size $-1$. In approximation the lateral motion is deterministic, up-steps move with velocity $-1$ and down-steps with velocity $1$. Steps annihilate at collisions. In addition pairs of adjacent up-/down-step are created according to a space-time Poisson process with 
uniform intensity, which for convenience will be set equal to $2$. 

For all these models, on a macroscopic scale the height is governed by a Hamilton-Jacobi equation of the form
\begin{equation}\label{1.5}
\partial_t h = \Phi(\partial_x h)\,,
\end{equation}
which expresses that the local change in height depends only on the local slope, $u = \partial_x h$. For the single-step model one finds $ \Phi(u)
= \tfrac{1}{2}(p-q) (1-u^2)$ and for the PNG model $ \Phi(u) = \sqrt{4+u^2}$. Some aspects of the  interface dynamics for the Ising model at low temperatures are discussed in 
\cite{Sp86}.

The statistical mechanics problem is to characterize the space-time fluctuations relative to the shape governed by \eqref{1.5}.
In general,  this turns out to be a challenging task and much of our understanding of the KPZ universality class relies on simplified models as 
single-step and PNG.   In 2000 K. Johansson \cite{Jo00}  studied the single-step model with 
$p = 1$ and wedge initial conditions, $h(j,0) = |j|$. He succeeded to determine the exact probability density function of $h(0,t)$ for large $t$, which constituted the starting point in the search for further integrable stochastic interface models and their universal properties.

\section{Scaling properties,\\ KPZ equation as  weak drive limit}
\label{sec2}
At first sight the KPZ equation seems to unrelated to the single-step model. To elucidate the connection we first study the scaling properties
of the KPZ equation and will use them to guess the limit in which the single-step model  is well approximated by the KPZ equation.
But before we note that, by rescaling $x,t,h$, any value of the material coefficients $\lambda,\nu,D$ can be achieved.
Also flipping $h$ to $-h$ is equivalent to flipping $\lambda$ to $-\lambda$.
Thus without loss of generality we set $\nu = \tfrac{1}{2}$, $D = 1$, which makes the formulas less clumsy. Hence the KPZ equation reads
\begin{equation}\label{2.0}
\partial_t h = \tfrac{1}{2} \lambda (\partial_x h)^2 + \tfrac{1}{2}\partial_x^2 h  +  \xi \,,
\end{equation}
keeping the dependence on the nonlinearity strength parameter $\lambda$.

Physically we are interested in the large space, long time behavior of the KPZ equation, with the view that in this limit the microscopic details will be irrelevant.  
First we note one important building block. The linear equation, $\lambda =0$, is a Gaussian process and its time-stationary measure is easily computed to be
given by
\begin{equation}\label{2.1}
Z^{-1}  \exp\big[ -\tfrac{1}{2}\int dx \big((\partial_x h(x))^2 -2 \mu \partial_x h(x)\big)\big] \,,
\end{equation}
where $\mu$ is the average slope. As a general experience, the nonlinear part of the drift will modify the time-stationary measure. 
However, the KPZ equation  \eqref{1.1a}  is very special 
to have the time-stationary measure independent of $\lambda$. One only has to observe that under the evolution governed by
\begin{equation}\label{1.1b}
\partial_t h = \tfrac{1}{2}\lambda (\partial_x h)^2 
\end{equation}
the time change of the action is
\begin{equation}\label{2.2}
 \frac{d}{dt} \int dx(\partial_x h(x))^2  =  2 \int dx\partial_x h(x)\partial_x \partial_t h(x) 
 =  \lambda \int dx \partial_x h(x) \partial_x  (\partial_x h(x))^2 =0\,.
\end{equation}
In principle one should worry also about the Jacobian. But formally the vector field in \eqref{1.1b} is divergence free and the Jacobian equals $1$. Our argument fails in higher dimensions. The steady state of the KPZ equation \eqref{1.1} is not known. 

We now transform to large scales by
\begin{equation}\label{2.3}
x  \leadsto  \epsilon^{-1}x ,  \qquad t  \leadsto  \epsilon^{-z}t \,.
\end{equation}
$\epsilon$ is the dimensionless scale parameter, $\epsilon \ll 1$, and $x,t$ on the right side are independent of $\epsilon$. $z$ is the dynamical scaling exponent,
which still has to be determined. The height field is  transformed to
\begin{equation}\label{2.4}
h_\epsilon(x,t) = \epsilon^{b}h(\epsilon^{-1}x , \epsilon^{-z}t) 
\end{equation}
with $b$ the fluctuation exponent. Recall that white noise satisfies $ (a_1a_2)^{1/2} \xi(a_1x,a_2t) = \xi(x,t)$. Thus inserting \eqref{2.4} in
\eqref{1.1a} one arrives at
\begin{equation}\label{2.5}
\partial_t h_\epsilon = \epsilon^{2-z-b}\tfrac{1}{2}\lambda (\partial_x h_\epsilon)^2  +\epsilon^{2-z} \tfrac{1}{2}\partial_x^2 h_\epsilon +\epsilon^{b + (1-z)/2} \xi\,.
\end{equation}
Since the   time-stationary measure is given by  Eq. \eqref{2.1}, the fluctuation exponent equals
 \begin{equation}\label{2.6}
b=\tfrac{1}{2}\,.
\end{equation}
To have in \eqref{2.5} the nonlinearity maintained implies then the dynamic exponent
\begin{equation}\label{2.7}
z =\tfrac{3}{2}\,.
\end{equation}

The KPZ equation has two well-separated and distinct noise scales. Locally the dynamics tries to maintain stationarity,
\textit{i.e.} $ x \mapsto h(x,t)$ has the statistical properties of a Brownian motion at some constant drift, in other words some locally averaged slope, which is constant on small scales but
still changing on coarser space-time scales. For large scales the nonlinearity dominates, but the evolution is still noisy. Its properties will have  to be computed. Scaling by itself is certainly not enough. But the gross features can be guessed already from Eq. \eqref{2.5}, setting $b = 1/2$ and $z = 3/2$. Then the ratios
 \textit{height} : \textit{space} : \textit{time} are given by 
$\epsilon^{-1/2}$ : $\epsilon^{-1}$ : $\epsilon^{-3/2}$. Choosing $\epsilon^{-3/2}$ as time unit, then, for large $t$, the typical height fluctuations are of order 
$t^{1/3}$ and correlations in $x$ are of order  $t^{2/3}$. Put differently, if one chooses a reference point $x_0$ and $|x - x_0| \ll t^{2/3}$,
then on that spatial interval the KPZ solution $ x \mapsto h(x,t)$ is like a Brownian motion with constant drift. To understand the statistical properties
for $|x - x_0| \simeq t^{2/3}$ requires further input.

With this background we can tackle the approximation through the single-step model. First note that if in  \eqref{2.0}  the nonlinearity    $\lambda$ is assumed to be equal to $\epsilon^{1/2}$ and
space is scaled to $\epsilon^{-1}x$, time  to $\epsilon^{-2}t$, then the scaled height function, $h_\epsilon(x,t) = \epsilon^{1/2}h(\epsilon^{-1}x , \epsilon^{-2}t)$, satisfies
\begin{equation}\label{2.8}
\partial_t h_\epsilon = \tfrac{1}{2}  (\partial_x h_\epsilon)^2  + \tfrac{1}{2}\partial_x^2 h_\epsilon + \xi\,.
\end{equation}
Thus the small nonlinearity is precisely balanced by  large space-time. 

Such a limit is meaningful also for the single-step model. To distinguish,
the single-step height is denoted by $h^\mathrm{step}(j,t)$. We adopt a lattice spacing $\epsilon$,
\textit{i.e.} $h^\mathrm{step}_\epsilon(x) = h^\mathrm{step}(\lfloor \epsilon^{-1} x \rfloor )$ with $\lfloor  \cdot \rfloor$ denoting integer 
part and $x$ independent of $\epsilon$. From our experience with much simpler equations we 
expect that in the limit of zero lattice spacing, with an appropriate simultaneous rescaling of $h^\mathrm{step}$ and $t$, one obtains some continuum equation. Our argument above suggests to choose a weak asymmetry as $p = \tfrac{1}{2}(1 +\kappa\sqrt{ \epsilon})$, $q = \tfrac{1}{2}(1 -\kappa\sqrt{ \epsilon})$, $\kappa > 0$. With this choice the rescaled height is
\begin{equation}\label{2.9a}
h^\mathrm{step}_\epsilon(x,t) = \epsilon^{1/2}h^\mathrm{step}(\lfloor\epsilon^{-1}x\rfloor , \epsilon^{-2}t). 
\end{equation}
Indeed there is a theorem by L. Bertini and G. Giacomin \cite{BeGi97}, which states that
\begin{equation}\label{2.9}
\lim_{\epsilon \to 0}\big(h^\mathrm{step}_\epsilon(x,t)-\epsilon^{-1}  \kappa t\big) = h(x,\kappa t)\,,
\end{equation}
where the right hand side is the solution to the KPZ equation \eqref{2.0} with $\lambda = 1$. In essence,  required is only that the initial height profile grows less than linearly at infinity.

The proof of the limit \eqref{2.9} is not at all obvious.  In the case of the single-step
model one relies on a transformation discovered by J. G\"{a}rtner \cite{Ga85}, which shifts the nonlinearity into the noise term. In fact, currently  there are only a few models for which such a limit can be established.
A major advance is the solution theory of M. Hairer \cite{Ha13}. In a related undertaking \cite{GuPe15}
the solution theory serves as a tool to prove that a discretized version of the KPZ equation, as proposed in \cite{SaSp09}, converges to the continuum equation \eqref{2.0}.

In our previous discussion we did not anticipate that, according to \eqref{2.9}, one has to switch to a moving frame of reference, whose velocity diverges as $\epsilon^{-1}$ for
$\epsilon \to 0$. In retrospect one might have expected. When approximating a continuum theory by a lattice based model, counter terms have to be subtracted.
In our case there is just a single term, independent of  $x$, which  is in spirit very similar to an energy renormalization in quantum field theory.

To summarize, the KPZ equation becomes exact in the limit of weak asymmetry. In the specific case of the Glauber model weak asymmetry  corresponds to a small magnetic field $h$ and a simultaneous rescaling of space-time and height. In this respect the KPZ
equation  is similar to other effective equations based on the availability of a small parameter. More exceptional is the feature to have a nonlinear limit dynamics which is still noisy.

\section{Eden type growth models}
\label{sec3}
Before proceeding to the analysis of the KPZ equation we discuss another class of growth processes, known as Eden models \cite{Ed61,BaSt95,Kr97,Me98}. This time the reason is not beautiful mathematics. Rather Eden models are currently the best
laboratory realized systems.
In the Eden model the ambient metastable phase is ignored entirely, in accord with single-step and PNG. Usually one starts from a seed and provides a rule for  potential growth sites. In a single update one of the growth sites is filled according to a uniform probability. To have an example with $\mathbb{Z}^2$ as underlying lattice, the origin is taken to be the seed. Given the connected cluster at time $t$, any site with distance 1 is a growth site. The cluster at time $t+1$ is obtained by filling one of the growth sites at random. After a long time a deterministic shape emerges, which looks circular, but nevertheless is anisotropic because of the underlying lattice.   For us the shape fluctuations are the main interest. They are described approximately by the
KPZ equation, but only in a small segment, since the KPZ height is the graph of a function.

The anisotropy of the Eden model implies that the coefficients of the approximating KPZ equation depend on the particular radial direction. This makes numerical simulations more difficult, since angular averaging will distort the universal result and is hence not advisable.  An isotropic Eden model would be preferred. One possibility is to have growth on $\mathbb{R}^2$,
where the basic building blocks are disks of fixed diameter. The seed is a disk located at the origin. Growth sites are disk centers such that the corresponding disk touches the current cluster, avoiding however any overlap with disks already present. 
The subsequent disk is attached according to the normalized Lebesgue measure on the union of arcs formed by the growth sites. 
By construction the limit shape is now a circle. In a  more physical variant, to every disk currently present a further disk is  attached, independently of all other disks, uniformly over all touching points, at constant rate, and subject to the constraint of no overlap \cite{Ta12}. 

A further variant of Eden type models is ballistic deposition. Along random rays orthogonal to the substrate, mass is transported and attached to the current cluster according to some prescribed rule. Early experiments on KPZ growth tried to realize such ballistic deposition.  Unfortunately it is difficult to control what precisely happens when a particle touches the growing surface. The more elegant realization is to have only a change of type at the interface. 
For example, in the smouldering paper experiment \cite{Hepaper}, the paper switches from unburned to burned at the flame front. 
No mass is transported. For $2+1$ dimensions, ballistic deposition has been revived by noting that larger molecules are more 
favorable building blocks \cite{Al14,HaPa14}.
\begin{figure}
\centering
\includegraphics[width=0.7\textwidth]{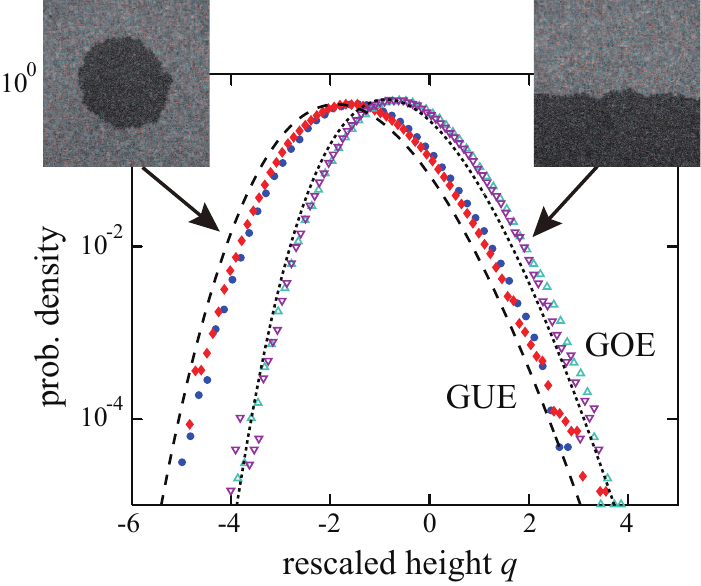}
\caption{
Histogram of the rescaled local height  $q$ for the circular (solid symbols) and flat (open symbols) interfaces. The blue and red solid symbols display the histograms for the circular interfaces at $t$ = 10 s and 30 s, while the turquoise and purple open symbols are for the flat interfaces at $t$ = 20 s and 60 s. The dashed and dotted curves show the densities of $\xi_\mathrm{cur}$ and $\xi_\mathrm{fla}$, respectively, see \eqref{1.14}, \eqref{1.12}. Note that the first moment is still shifting, while the higher cumulants have settled already to their asymptotic values.}
\label{fig1}
\end{figure}

K. Takeuchi and M. Sano \cite{TaSa12} had the ingenious idea to use turbulent liquid crystal for realizing a stable-metastable interface. In the actual experiment the liquid crystal film is $16\times16$ mm at a height of 12 $\mu$m. The rod-like molecules are on average aligned to be orthogonal to the confining plates. Hence the system is in-plane isotropic. There is an external electric field, uniform in space and oscillating in time, which makes the bulk phases turbulent, thus ensuring rapid relaxation. In fact these are nonequilibrium steady states. One carefully selects a point in the phase diagram, at which a stable (DSM2) and a metastable (DSM1) phase coexist. The two phases are easily distinguished through transmission of light. DSM2 is black, transmitting no light, while DSM1 appears in a grey color.  The film is prepared in the metastable DSM1 phase and a seed of DSM2 is planted by a very sharp laser pointer. Alternatively the laser may print a line seed, through which the dependence on initial conditions can be studied.  The cluster grows to its maximal size in approximately 40 - 60 s. On the order of $5\times 10^3$ repeats are carried out. Adding the angular directions, one achieves a large sample set. In Fig. \ref{fig1} we show the histogram, on a logarithmic scale, for a point seed and a line seed.  The respective theoretical predictions are discussed in Sect. \ref{sec4}.
In fact, the transition from metastable to stable seems to be a complicated physical process and 
currently there is little understanding of the precise mechanism. But empirically the isotropic Eden model captures the main 
features of the growing interface. For further details we refer to \cite{TaSa12}.

\section{The KPZ universality class}
\label{sec4}
Starting in 2000 for a variety of growth models exact universal scaling properties have been obtained,
including the KPZ equation itself. Some of the most important results will be listed. To disentangle however which result has been proved for which model is beyond the present scope and has to be looked up in more specialized articles \cite{FeSp11,BoGo12,Ba14}. In particular I recommend the review article \cite{CorwinRev} by I. Corwin, which covers the field up to 2011. For better readability, I list the results as if obtained for the KPZ equation. In some cases this is actually correct \cite{AmCoQu11,SaSp11,BoCoFeV14}. In other cases the corresponding solution of the KPZ equation is widely open, but the asymptotics has been obtained using another model in the KPZ universality class. We are still at the stage at which very specific models are analysed in considerable detail. To recall, the KPZ equation  is written as
\begin{equation}\label{1.10}
\partial_t h = \tfrac{1}{2} \lambda (\partial_x h)^2 + \tfrac{1}{2}\partial_x^2 h  +  \xi \,.
\end{equation}
(i) \textit{Initial conditions}. The long time asymptotics of the solution depends on the initial conditions. Three standard classes have been identified. They are (IC1) \textit{flat}, $h(x,0) = 0$; (IC2) \textit{curved}, \textit{e.g.} $h(x,0) = - x^2/2 $; (IC3) \textit{stationary}, $x \mapsto h(x,0) = B(x)$, 
where $B(x)$ is a two-sided Brownian motion pinned as $B(0) = 0$. Note that also $h(x,0) = ux + B(x)$ is stationary, but 
$\langle B(x)^2\rangle = |x|$ is required by our choice of parameters.
One can also consider domain walls formed by of such initial data. E.g. $h(x,0) = 0$ for $x \leq 0$ and $x \mapsto h(x,0)$ a Brownian motion for $x \geq 0$. If the focus is far to the left, then one is in class (IC1) and far to the right in class (IC3). But near $x =0$ novel cross-over statistical properties will be realized \cite{CorwinRev}.\medskip\\
(ii) \textit{Observables}. In statistical physics one learns that correlations, possibly higher order cumulants,
are the central goal. KPZ is actually an area where full probability density functions are of considerable advantage. 
They  seem to characterize more sharply the KPZ universality class  than scaling exponents. In numerical simulations, and in  experiments,  the line shape  often settles earlier than a definite power law.\medskip\\
\textit{One-point distributions}. The most basic observable is the long time statistics of the height at one spatial reference point, which for the initial conditions from above can be taken as $x=0$.  As a generic result,
\begin{equation}\label{1.11}
h(0,t) \simeq c_\diamond t +\sigma_\diamond (\Gamma_\diamond t)^{1/3} \xi_\diamond\,,\qquad \sigma_\diamond =\pm 1\,,
\end{equation}
valid for long times. The subscript ${}_\diamond$ stands for  either ``fla'', or ``cur'', or ``sta'', depending on the initial conditions. If one samples $h(0,t)$ at some large time $t$, then the distribution is shifted by $c_\diamond t$ and the fluctuations are of order  $t^{1/3}$ with a random amplitude characterized by  the random variable $\xi_\diamond$.  The nonuniversal factor of the fluctuating term is best collected through the rate $\Gamma_\diamond>0$. $\xi_\diamond$ is defined with a particular sign convention. But the random amplitude could be either $\xi_\diamond$ or $-\xi_\diamond$,
depending on the particular situation.
 In Sect. \ref{sec2} the exponent $\tfrac{1}{3}$ has been anticipated already on the basis of a simple scaling argument. But now we assert
in addition the full probability density function. 
 $c_\diamond$ and $\Gamma_\diamond$ are coefficients depending on the model. All other features are universal. 
 For the KPZ equation one obtains $c_\diamond = - \frac{1}{24}\lambda^3 t$, $\sigma_\diamond = \mathrm{sgn}(\lambda)$, $\Gamma_\mathrm{fla} = \tfrac{1}{8}|\lambda|$, $\Gamma_\mathrm{cur} = \tfrac{1}{2}|\lambda|$, $\Gamma_\mathrm{sta} = \tfrac{1}{2}|\lambda|$. The nonuniversal coefficients are known  also for a few other models. $c_\diamond$, $\sigma_\diamond$, and $\Gamma_\diamond$, 
 may take different values in distinct equations.
 
For (IC1) the random amplitude $\xi_\mathrm{fla}$ is distributed as GOE Tracy-Widom, for (IC2) the amplitude $\xi_\mathrm{cur}$  is distributed as GUE Tracy-Widom, and for (IC3) the amplitude
$\xi_\mathrm{sta}$  is distributed as Baik-Rains. These are non-Gaussian random variables and their distribution functions are  written in terms of  Fredholm determinants. More explicitly,
for GOE Tracy-Widom \cite{TW93}
\begin{equation}\label{1.12}
\mathbb{P}(\xi_\mathrm{fla} \leq s)= \det(1- K_{1,s})_{L^2(\mathbb{R}_+)} = F_1(s)\,,
\end{equation}
where
\begin{equation}\label{1.13}
K_{1,s} (x,y)=\mathrm{Ai}(x+y +s)
\end{equation}
with $\mathrm{Ai}$ the standard Airy function, see \cite{FeSp07} for this particular representation. 
For GUE Tracy-Widom \cite{TW93}
\begin{equation}\label{1.14}
\mathbb{P}(\xi_\mathrm{cur} \leq s) = \det(1- K_{2,s})_{L^2(\mathbb{R}_+)}= F_2(s)\,,
\end{equation}
where
\begin{equation}\label{1.15}
K_{2,s} (x,y) = \int_0^\infty du \mathrm{Ai}(x+u +s)\mathrm{Ai}(y+ u +s)\,.
\end{equation}
The Baik-Rains \cite{BaRa00} distribution has a more complicated expression,
\begin{equation}\label{1.16}
\mathbb{P}(\xi_\mathrm{sta} \leq s)=  F_0(s) =\frac{d}{ds}\big(g(s) F_2(s)\big)
\end{equation}
with 
\begin{equation}\label{1.17}
g(s) = s + \langle 1, (1 - K_{2,s} )^{-1}(K_{1,s} -K_{2,s} )1\rangle_{L^2(\mathbb{R}_+)}\,.
\end{equation}
All determinants are on the Hilbert space $L^2(\mathbb{R}_+)$ with inner product 
$\langle \cdot, \cdot\rangle_{L^2(\mathbb{R}_+)}$ and $1$ is the constant function.

$F_1$ and   $F_2$ have appeared before in the context of random matrix theory, 
where they characterize the fluctuations of the largest eigenvalue of GOE and GUE random matrices. The Baik-Rains distribution does not seem to have an obvious connection to random matrix theory. 

Early numerical plots were based on the connection to the Hastings-McLeod solution of the Painlev\'{e} II
differential equation. Since this solution is unstable, one has to employ an  ultra-precise shooting algorithm. F. Bornemann
\cite{Bo09} pointed out that a direct numerical evaluation of the suitably approximated Fredholm determinant 
is a more accessible approach and works equally well in cases when no connection to a differential equation
is available. 

Instead of the reference point $x=0$ one can also consider $x$ along the ray $\{x=vt\}$. Then, for long times,
\begin{equation}\label{1.18}
h(vt,t) \simeq c_\diamond(v) t + \sigma_\diamond (\Gamma_\diamond(v) t)^{1/3} \xi_\diamond\,.
\end{equation}
For flat  initial conditions, $h(vt,t)$ is independent of $v$ by translation invariance. For curved initial conditions the velocity $v$ is arbitrary, within limits set by the model, but $c_\mathrm{cur},\Gamma_\mathrm{cur}$ depend on $v$, in general. However in the stationary case, there is only one specific 
velocity, $v_0$, for which anomalous fluctuations of order $t^{1/3}$ are observed. For all other rays the fluctuations are  Gaussian
of size $\sqrt{t}$. Physically, $v_0$ is the propagation velocity 
of a small localized perturbation in the slope. For the KPZ equation $v_0 = 0$ and the asymptotics \eqref{1.11} holds also for $\diamond = \mathrm{sta}$. 
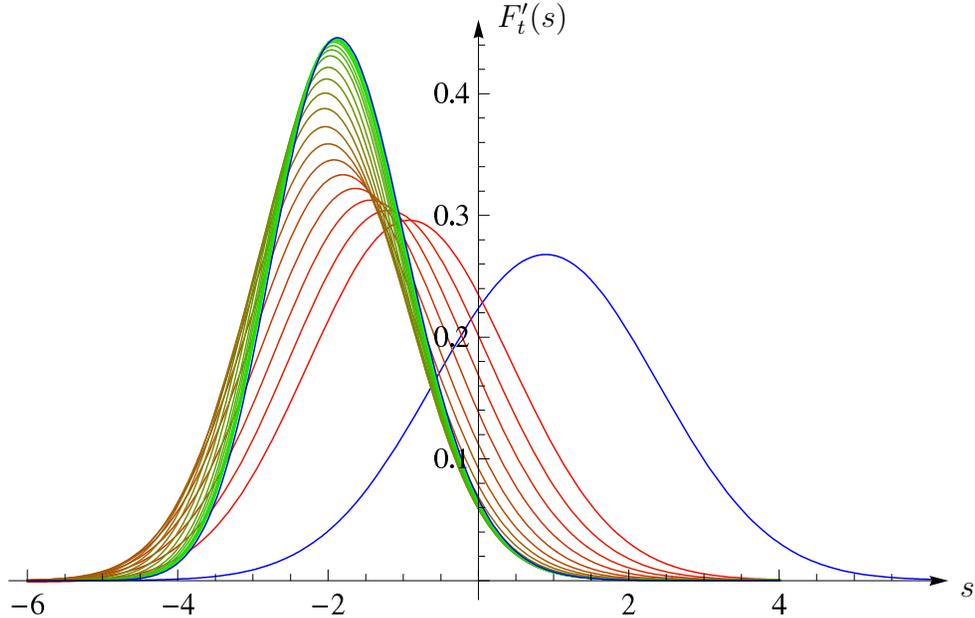
\begin{figure}
\centering
\begin{center}\begin{picture}(0,0)\put(185,223){$F_{t}'(s)$}\put(360,7){$s$}\end{picture}\includegraphics[width=125mm]{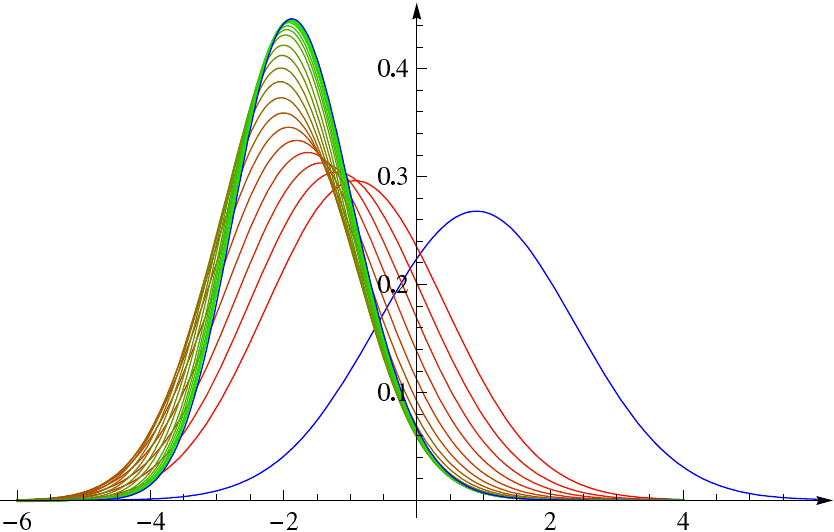}\end{center}
\caption{Probability density function $F_{t}'(s)$ for time $t$ from short times (red, lower curves) to long times (green, upper curves) for  $t$ ranging from 0.25 to 20,000, see \cite{PrSp12} for further details.  For $t\to 0$, $F_{t}'(s)$ becomes a Gaussian (rightmost blue curve) and for $t\to\infty$, the density converges to the GUE Tracy-Widom distribution $F'_2(s)$ (upper blue curve). Note that, in contrast to the experiment, the long time limit is approached from the left.}
\label{fig2}
\end{figure}

A particular case of curved initial conditions is the KPZ equation with sharp wedge initial data, \textit{i.e.} $h(x,0) = \lim_{\delta \to 0} - \delta^{-1} |x| - \log (2\delta)$. Then $h(x,t) = - (x^2/2 \lambda t) -\tfrac{1}{24} \lambda^3 t +\eta(x,t)$ and
$x \mapsto \eta(x,t)$ 
is stationary for fixed $t$. The exact  probability density function of $\eta(0,t)$, denoted by  $F'_t(s)$, can be written 
 in terms of the difference of two Fredholm determinants \cite{AmCoQu11,SaSp11}.  In Fig. \ref{fig2} we show a  time sequence of such densities \cite{PrSp12}, obtained using  the Bornemann method. At early times the density is Gaussian with variance of order $t^{1/4}$, which then crosses over to the GUE Tracy-Widom density on scale $t^{1/3}$.
\medskip\\
\textit{Multi-point distributions}. Instead of the single reference point $(vt,t)$, the joint distribution of the height at several space-time points could be considered. At such generality not much is known. For curved initial data, two times at the same space
point, \textit{e.g.} the joint distribution of $h(0,t),h(0,2t)$, are recently studied by K. Johansson \cite{Jo14}. But otherwise all
results refer to a single time, but with an arbitrary number of spatial points.  As anticipated in Sect. \ref{sec2}, to obtain a nondegenerate universal limit the space points must be separated on the scale $t^{2/3}$. The asymptotics \eqref{1.11} generalizes to 
\begin{equation}\label{1.19}
h(wt^{2/3},t) \simeq c_\diamond t + \sigma_\diamond(\Gamma_\diamond t)^{1/3} \mathcal{A}_\diamond (w)\,,
\end{equation}
as a stochastic process in $w$, which means that the finite dimensional distributions from the left, \textit{i.e.} joint distributions for a finite number of reference points, converge to the one on the right. The limit process $\mathcal{A}_\diamond (w)$ is known as Airy process. For each of the three classes of initial conditions there is a distinct Airy process.\medskip\\
\textit{Stationary covariance}. With two-sided Brownian motion as initial conditions the height field is not stationary in the usual sense of the word. Rather, if $h(x,0) = B(x)$, then
$x \mapsto h(x,t) - h(0,t)$ is again two-sided Brownian motion. Note that the random shift $h(0,t) - h(0,0)$ is correlated with   $h(x,t) - h(0,t)$. Stationarity in the conventional sense is achieved by considering instead the slope $u(x,t) =\partial_x h(x,t)$, which is governed by the stochastic Burgers equation
\begin{equation}\label{1.20}
\partial_t u -\partial_x\big(\tfrac{1}{2}\lambda u^2 + \tfrac{1}{2} \partial_x u +  \xi\big) =0\,.
\end{equation}
In this case,  the time-stationary measure is unit strength white noise in $x$ and, for such initial conditions,  $u(x,t)$ is a random field stationary in both space and time. 

For the linear case, $\lambda =0$, the covariance is easily computed to be given by
\begin{equation}\label{1.21}
\langle u(x,t) u(x',t')\rangle =  (2\pi  |t - t'|)^{-1/2} \exp\big[ - (x-x')^2/2|t-t'|\big]\,.
\end{equation}
A  small perturbation in the slope spreads diffusively. For $\lambda \neq 0$ one has to rely on the asymptotics in \eqref{1.19} with $\diamond = \mathrm{sta}$.
One first notes that
\begin{equation}\label{1.22}
\partial_x^2 \big\langle \big(h(x,t) -h(0,0) - c_\mathrm{sta} t \big)^2\big\rangle = 2 \langle u(x,t) u(0,0)\rangle
\end{equation}
and,  using \eqref{1.19}, infers that
\begin{equation}\label{1.23}
\big\langle \big(h(wt^{2/3},t) - c_\mathrm{sta} t \big)^2\big\rangle  =\big\langle \big((\Gamma_\mathrm{sta} t)^{1/3} \mathcal{A}_\mathrm{sta} (w) 
\big)^2\big\rangle
\end{equation}
valid for large $t$. There is an explicit formula for the probability distribution of $\mathcal{A}_\mathrm{sta} (w) $. Hence, one has to compute its second moment and twice differentiate w.r.t.  $w$ to obtain the universal stationary scaling function.   The result is a self-similar two-point function 
\begin{equation}\label{1.24}
\langle u(x,t)u(0,0)\rangle \simeq (\Gamma_\mathrm{sta} |t|)^{-2/3}f_\mathrm{KPZ}((\Gamma_\mathrm{sta} |t|)^{-2/3}x)\,,\quad \Gamma_\mathrm{sta} = \sqrt{2}|\lambda|\,,
\end{equation}
valid for large $|x|,|t|$. The function $f_\mathrm{KPZ}$   is tabulated in~\cite{Prhp}, denoted there by $f$. Its properties are $f_{\mathrm{KPZ}}\geq 0$, $\int_\mathbb{R} dxf_{\mathrm{KPZ}}(x)=1$, $f_{\mathrm{KPZ}}(x)=f_{\mathrm{KPZ}}(-x)$, $\int_\mathbb{R} dxf_{\mathrm{KPZ}}(x)x^2 \simeq 0.510523$. $f_{\mathrm{KPZ}}$ looks like a Gaussian with a large $|x|$ decay as $\exp[-0.295|x|^{3}]$. Plots are provided in~\cite{PrSp04,Prhp}.
As required by the conservation law,  $\int_\mathbb{R} dx \langle u(x,t)u(0,0)\rangle =  1$.

 \section{Directed polymers in a random medium}
\label{sec5a}
One can rewrite the KPZ equation as a problem in equilibrium statistical mechanics of
disordered systems. This step is a one-to-one map, no information is lost, and offers a different intuition on the
KPZ equation. Besides, new tools become available. As noted already by Hopf \cite{Hopf} and Cole \cite{Cole} for the dissipative Burgers equation, \eqref{1.20} with zero noise, the transformation
\begin{equation}\label{5a.1}
Z(x,t) = \mathrm{e}^{\lambda h(x,t)}
\end{equation}
``linearizes'' the KPZ equation as
\begin{equation}\label{5a.2}
\partial_t Z(x,t) =  \tfrac{1}{2} \partial_x^2Z(x,t)  +\lambda\xi(x,t)Z(x,t)\,.
\end{equation}
Eq. \eqref{5a.2} is the heat equation with a space-time random potential, hence also called stochastic heat equation. Following
Feynman and Kac, its solution can be written as the expectation over an auxiliary Brownian motion, $b(t)$,
\begin{equation}\label{5a.3}
Z(x,t) =  \mathbb{E}_{(x,t) }  \Big(  \mathrm{exp}\big( \lambda  \int_0^t ds  \,\xi(b(s),s) \big) Z_0(b(t)) \Big)\,.
\end{equation}
$b(t)$ is the directed polymer, which starts at $(x,t)$ and moves backwards in time to end at $(b(t),0)$. The directed polymer 
has an intrinsic elastic energy, implicit in  the expectation $\mathbb{E}_{(x,t)}$, and a potential energy obtained by integrating the random potential
$\xi(x,s)$ along its path.
$Z_0(x) = \mathrm{e}^{\lambda h(x,0)}$ is the initial condition. The partition function $Z(x,t)$ is the sum over all paths weighted with the Boltzmann factor.
Since the potential energy is random, the partition function is random and we arrived at a problem from the theory of disordered systems, which studies systems in thermal equilibrium for which the coupling constants appearing in the energy are random, but regarded as fixed for thermal averages. At the end, our interest is the
random free energy
\begin{equation}\label{5a.4}
h(x,t) = \lambda^{-1}\log Z(x,t)\,.
\end{equation}
It has a leading term linear in $t$, which is self-averaging, \textit{i.e.}
\begin{equation}\label{5a.5}
\lim_{t \to \infty} t^{-1} \lambda^{-1}\log Z(x,t) = v_0 
\end{equation}
almost surely. For growing interfaces the  key point are the fluctuations of the free energy, a not so well studied quantity for disordered systems.

The directed polymer in  \eqref{5a.3} is called a continuum directed polymer. Since we are interested in large scale properties,
 its local properties can be modelled fairly freely. A popular, and natural, choice is to replace the space-time continuum by the discrete lattice $\mathbb{Z}^{2}$ and the directed polymer by an up-right path $ \omega $, starting at $(0,0)$ and ending at $(N,N)$, say.
The white noise is replaced by independent identically distributed random variables $ \xi_{i,j}$, $(i,j) \in  \mathbb{Z}^{2}$. The energy of a $2N$-step directed polymer is now
\begin{equation}\label{5a.6} 
E(\omega) = \sum _{\ell= 1}^{2N+1} \xi_{\omega(\ell)}
\end{equation}
and the discretized version of the partition function \eqref{5a.3} reads
\begin{equation}\label{5a.7} 
Z_\beta(N,N) =\sum_ {\omega: (0,0) \to (N,N)}\mathrm{e}^{- \beta E(\omega)} \,,
\end{equation}
where we introduced the more conventional inverse temperature $\beta$ as parameter. In analogy, the height function is defined by
\begin{equation}\label{5a.7a}
h_\beta(N,N) = -\beta^{-1}\log Z_\beta (N,N)\,.
\end{equation}
This problem is called a point-to-point directed polymer and corresponds to a sharp wedge
 in the language of height functions. On the other hand, for flat initial conditions, $h(0,x) =0$,
 implying $Z_0(x) = 1$, which corresponds to sum over all directed polymers with only one end point fixed. This is called point-to-line directed polymer. From this perspective it is less surprising that flat and curved initial conditions have distinct fluctuation behavior. 
 
 In a discretized version, as in  \eqref{5a.7}, one can take the limit $\beta \to \infty$.  Then the log is traded against the exponential and the finite temperature problem turns into a ground state problem. Hence, the height function at a given point is
\begin{equation}\label{5a.8} 
h_\infty (N,N) = \min_ {\omega: (0,0) \to (N,N)} E(\omega)\,.
\end{equation} 
No surprise, one expects, and proves for a few very specific distributions of $\xi_{i,j}$, that
\begin{equation}\label{5a.9} 
h_\infty(N,N) \simeq c_\mathrm{DP} N + (\Gamma_\mathrm{DP} N)^{1/3} \xi_\mathrm{GUE}
\end{equation}  
for large $N$ \cite{Jo00,BoCoRe13,Se15}. The PNG model is also included in the list as a shot noise limit. The random medium is now a homogeneous, two-dimensional Poisson point process. An admissible path is continuous, increasing in both coordinates, and consists of linear segments bordered by Poisson points. Each Poisson point carries a negative unit of energy and, as before, one studies the ground state energy.
An optimal path is defined by transversing a maximal number of Poisson points.

In the representation through a directed polymer, one can ask how the continuum directed polymer is approximated through a discrete version. This is just like approximating the KPZ equation by a discrete growth model.
In fact, the continuum directed polymer is obtained through a weak noise limit. We  refer to  \cite{AlKhQu14}, where a proof under fairly general assumptions is carried through. 

\section{Replica solutions}
\label{sec5c}
As noted already early on  \cite{Ka87}, there are closed evolution equations for the moments of the partition function $Z(x,t)$ as defined in 
\eqref{5a.3}. Let us consider the
$n$-th moment, now written with $n$ auxiliary independent Brownian motions, called the replicas. Denoting the white noise average by $\langle \cdot
\rangle$, one arrives at
\begin{equation}\label{5c.1}
\big\langle \prod_{j=1}^n Z(x_j,t) \big\rangle =
\big\langle \prod_{j=1}^n\mathbb{E}_{(x_j,t)} \big(\mathrm{e}^{\lambda\int_0^t ds \xi(b_j(s),s)} Z_0(b_j(t))\big)\big\rangle\,.
\end{equation}
The white noise average can be carried out explicitly and is given by the exponential of 
\begin{equation}\label{5c.2}
\tfrac{1}{2}\lambda^2\sum_{i,j=1}^n \int_0^t \int_0^tds_1ds_2 \,\delta(s_1-s_2) \,\delta(b_i(s_1) -b_j(s_2))\,.
\end{equation}
The summand with $i=j$ is defined only when smearing the $\delta$-function. However, if for the stochastic integral in Eq. \eqref{5a.3} one adopts the  It\^{o} discretization, then  the diagonal term $i = j$
has to be omitted. The double time integration reduces trivially to a single one. Hence, using the Feynman-Kac formula backwards, one obtains
\begin{equation}\label{5c.3}
\big\langle \prod_{j=1}^n Z(x_j,t) \big\rangle = \langle x_1,...,x_n|\mathrm{e}^{-H_nt}|(Z_0)^{\otimes n}\rangle\,.
\end{equation}
Here $H_n$ is the $n$-particle Lieb-Liniger quantum Hamiltonian on the real line with attractive $\delta$-interaction,
\begin{equation}\label{5c.4}
H_n = - \tfrac{1}{2} \sum_{j=1}^n \partial_{x_j}^2 - \tfrac{1}{2} \lambda^2\sum_{i \neq j = 1}^n\delta(x_i - x_j)\,.
\end{equation}
The quantum propagator acts on the initial product wave function $\prod_{j= 1}^n Z_0(x_j)$, denoted by $(Z_0)^{\otimes n}$,
and is evaluated at the point $(x_1,...,x_n)$. Since $(Z_0)^{\otimes n}$ is symmetric,
only the restriction of $\exp [-H_n t]$   to the subspace of permutation symmetric wave functions, the bosonic subspace, in $L^2(\mathbb{R}^n)$ is required.
As a result the right hand side of (\ref{5c.3}) is a symmetric function, as it should be.

For curved initial data, in the sharp wedge approximation $Z_0(x) = \delta(x)$, one obtains
\begin{equation}\label{5c.5}
\langle Z_\mathrm{cur}(0,t)^n \rangle = \langle 0|\mathrm{e}^{-H_n t}|0\rangle
\end{equation}
with shorthand $|0\rangle = |0,...,0\rangle$. For flat initial conditions, $h(x,0) = 0$, the $n$-th moment of the partition function reads
\begin{equation}\label{5c.6}
\langle Z_\mathrm{fla}(0,t)^n \rangle = \int_{\mathbb{R}^n}dx_1...dx_n\langle 0|\mathrm{e}^{-H_n t}|x_1,...,x_n\rangle\,.
\end{equation}
Also for stationary initial data there is a concise formula, the initial wave function however being no longer of product form,
\begin{equation}\label{5c.f}
\langle Z_\mathrm{sta}(0,t)^n \rangle = \int_{\mathbb{R}^n}dx_1...dx_n\langle 0|\mathrm{e}^{-H_n t}|x_1,...,x_n\rangle
\exp\big[\tfrac{1}{2}\big\langle\big(\sum_{j=1}^n B(x_j)\big)^2 \big\rangle\big]\,,
\end{equation}
 where the right average is over the two-sided Brownian motion $B(x)$.
 
Of course, the general hope is to extract from the moments some information on the distribution of $\log Z(x,t)$. Unfortunately the moments 
diverge as $\exp[n^3]$ and one is forced to  fall back on formal resummation procedures. Even then there are prior difficulties. Firstly from the Bethe ansatz solution of the Lieb-Liniger model, one has to deduce a sufficiently concise formula for the particular matrix element of the propagator. It is not known how to proceed for general initial data, but for the three canonical initial conditions this step has been accomplished,
at increasing complexity from wedge \cite{CLDR2010,Dotsenko2010,Do10}, to stationary \cite{ImSa12,ImSa13},
and to flat \cite{CLD2011,DC2012}. While for the curved and stationary  cases there are corresponding rigorous results, the flat initial conditions are yet to be resolved \cite{OQR14,OQR15}. The second difficulty is part of working with a badly divergent series.
One is not allowed to somehow cut, or otherwise approximate,  the series at intermediate steps.

The current results all deal with a single reference point. For the joint distribution at several space points, one relies on an intermediate decoupling assumption \cite{PrSp11a,PrSp11b}. On a large scale the resulting expressions agree with the corresponding ones from lattice models, indicating that the decoupling is valid at least in approximation.

\section{Statistical mechanics of line ensembles}
\label{sec5b}
There is a second mapping, which is more hidden and has been discovered only 15 years after the publication of the KPZ paper. In contrast to the Cole-Hopf transformation, the second mapping deals only with the data at some fixed time and thus provides partial information only. Nevertheless, it is this mapping through which many of the universal results were obtained first. Whether such mapping can be defined for the KPZ equation is not known at the moment and we turn instead to the PNG model \cite{PrSp03}.  We consider droplet growth, which means that there is an initial ground layer expanding linearly in time up to $[-t,t]$ and all nucleation events outside this layer are suppressed. At time $t$ we have 
the random height profile $x \mapsto h_\mathrm{PNG}(x,t)$. Now we claim that the statistics of $h_\mathrm{PNG}(x,t)$ at fixed $t$
can be obtained through a direct construction, completely avoiding an explicit solution of the dynamics. We choose  $x$ as running parameter, $|x|\leq t$, and consider independent, time-continuous, symmetric, simple random walks $\omega_n(x)$, $n = 0,-1,...$\,. $\omega_n$ is already conditioned on
$\omega_n(\pm t) = n$. Now we pick $M > 0$ and further condition on the event that the walks $\omega_n(x), n = 0,...,-M$, do not intersect.
Finally we take the limit $M \to \infty$. This limit exists, since there is some smallest random index $m$ such that $\omega_m(x) = m$
for all $x$.
The such conditioned non-intersecting random walks are denoted  again by $\omega_n(x)$. The theorem is that the distribution of 
$h_\mathrm{PNG}(x,t)$, $t$ fixed, is identical to the one of the top random walk $\omega_0(x)$. 

Because of entropic repulsion the typical shape of $\omega_0(x)$ is a droplet of the form of a semicircle, $h(x) = 2\sqrt{1 - x^2}$. The collection $\{\omega_n(x), n\in \mathbb{Z_-} \}$
is called a non-intersecting line ensemble, which this time is an object of equilibrium statistical mechanics. More generally, such an ensemble is 
defined through the Boltzmann weight
\begin{equation}\label{5b.1} 
Z^{-1 } \exp\Big[-\beta \sum_{n=-M}^{-1} \int_{-t}^t dyV(\omega_{n+1}(y)- \omega_ {n}(y))\Big]\,,   
\end{equation}   
where $V$ is a short range, strongly repulsive hard core potential.  \eqref{5b.1} defines a random field $\{\zeta(x,j), x\in [-t,t],
j\in \mathbb{Z}\}$, where $\zeta(x,j) = 1$ if a line passes through $(x,j)$ and $\zeta(x,j) = 0$ otherwise. Of course, one still has to specify the boundary conditions. 
In our example the lines are pinned at the border lines $\{ |x| = \pm t\} $. Thereby the equilibrium measure becomes inhomogeneous, both in $x$ and $j$.

Our mapping comes with an additional powerful tool. In the limit of an infinitely strong point repulsion, which is equivalent to the conditioning discussed above, the random walks $x \mapsto \omega_n(x)$ are the world lines of non-interacting fermions. $x$ is the Euclidean time and $j$ is space. The fermions start at $x = -t$ with the lattice $\mathbb{Z}$ half-filled from $- \infty$ to $0$. They evolve in imaginary time by the standard symmetric nearest neighbor hopping. Thus the one-particle Hamiltonian is the nearest neighbor Laplacian $- (\Delta f)_j = - f_{j+1} -f_{j-1} +2f_j$. At time $x =t$ the fermions have to return to their original positions.  Such problem can be handled by free fermion techniques. To study $ \omega_0(0)$ it is convenient to consider the full collection of points $\{\omega_n(0), n = 0,-1,... \}$. This turns out to be the 
ground state of free fermions subject to a linear external potential $j/t$. Just like in the case of a gravitational potential there is a top fermion. As $t \to\infty$, using that the slope of the linear potential decreases as $1/t$, the position of the top particle has the distribution
\begin{equation}\label{5b.2} 
\omega_0(0) = h_\mathrm{PNG}(0,t) \simeq 2t + (\Gamma_\mathrm{PNG} t)^{1/3} \xi_\mathrm{cur}\,,  
\end{equation}   
valid for large $t$. Extending to several reference points, one concludes convergence to the full
$\mathcal{A}_\mathrm{cur}(w)$ process. 

A more complete discussion can be found in my write-up for the 2005 Summer School on ``Fundamental Problems in 
Statistical Mechanics'' at Leuven  \cite{Sp06}.


\section{Noisy local conservation laws}
\label{sec6}
Any one of the topics mentioned so far deserves further explanations. But this would easily run oversize.
Instead I will focus in much greater detail on one aspect, which is a recent development and covers  physics 
yet different from growth processes.

On a purely formal level the generalization consists of replacing the scalar height $h(x,t)$ by an $n$-vector
$\vec {h} = (h_1,...,h_n)$ \cite{FeSS13,Sp14}. In our applications the relevant quantities will be the slopes 
$\partial_x \vec {h}$. Thus we start from the KPZ equation in the slope form \eqref{1.20}  and generalize it to
\begin{equation}\label{0.1}
\partial_t u_\alpha + \partial_x \big((A \vec {u}\,)_\alpha + \vec{u} \cdot (H^\alpha \vec {u}) +
(D \vec {u}\,)_\alpha + (B \vec {\xi}\,)_\alpha \big)  =0\, .
\end{equation}
These are $n$ coupled conservation laws, $u_\alpha$, $\alpha = 1,...,n$, being the conserved fields.    
The currents have three pieces: (i) \textit{a nonlinear current}. We included terms only up to quadratic  order, since by power counting higher orders 
are expected to be irrelevant.
The linear part is specified by the $n \times n$ matrix $A$. For $n=1$ it could be removed by switching to a  frame 
moving with constant velocity. But for larger $n$ this will not be possible unless all eigenvalues of $A$ coincide. The quadratic part is specified by the symmetric
Hessians $H^\alpha$. (ii) \textit{dissipation}. This term is proportional to the gradients. The diffusion matrix, $D$, has positive eigenvalues. 
In principle $D$ could depend also on $\vec {u}$, but this is regarded as a higher order effect.
(iii) \textit{fluctuating currents}. $\xi_\alpha$ is Gaussian white noise with independent components, $\langle \xi_\alpha(x,t) \rangle = 0$,
$\langle  \xi_{\alpha}(x,t) \xi_{\alpha'}(x',t')\rangle = \delta_{\alpha\alpha'} \delta(x - x')\delta(t - t')$. The matrix $B$ encodes possible correlations 
between the random currents. 

The generalization \eqref{0.1} may look more natural after providing a few examples. First we return to the single-step model. Its height has slopes $\pm1$.
We regard $-1$ as the position of a particle and $1$ as an empty site. Then under the single step dynamics the particles form a stochastic system
known as asymmetric simple exclusion process (ASEP). Particles hop on the lattice $\mathbb{Z}$. Independently they jump to the right with rate $p$ and to left
with rate $q$. An attempted jump is suppressed in case it would lead to a double occupancy  (the exclusion rule). ``Simple'' refers to nearest neighbor jumps only. Clearly, the particle number is the only conserved field. The steady states are Bernoulli. Therefore, at density $\rho$, the average current, $\mathsf{j}$,
equals $\mathsf{j}(\rho) = (p-q) \rho(1 - \rho)$. As explained in Sect. \ref{sec2}, on the mesoscopic scale the density is governed by the stochastic Burgers equation  \eqref{1.20}, at least for small $|p-q| \neq 0$.
To generalize from one to $n$ components we introduce particles of type $\alpha = 1,...,n$. $\alpha$ particles jump under the exclusion rule on lane
$\alpha$, but the jump rate depends on the occupancy of the corresponding sites on
the other lanes. In the limit of weak asymmetry this then leads to \eqref{0.1}. Obviously, our theme allows for many variations. One particular version is to have two components with particles hoppping on $\mathbb{Z}$,
subject to exclusion. If the exchange rates depend only on nearest neighbor occupations, one arrives at the well studied AHR model \cite{AHR99}.

Also the PNG model can viewed as stochastic motion of particles. Now there are two types, $\pm$, already. A $+$ particle is located at a down-step
moving to the right and  a $-$ particle is located at an up-step
moving to the left, both at unit speed. Particles annihilate at collisions. In addition,  point-like $-|+$ pairs are generated at constant rate. We denote the two 
components by $\rho_ +$, $\rho_-$. Only  $\rho = \rho_+ - \rho_-$ is conserved, while $\mathsf{j} =\rho_+ + \rho_-$ is not conserved.  As anticipated by notation, $\mathsf{j}$ is the current for $\rho$. Using that the stationary states are uniform Poisson, one concludes that the current-density relation is $\mathsf{j}(\rho) = \sqrt{4 + \rho^2}$. However, the PNG model has no natural asymmetry parameter. So there is
no appropriate limit leading to the stochastic Burgers equation. Of course, multi-component versions are still easily invented.

Even at that general level there is already a crucial distinction. The dominant term in \eqref{0.1} is the linear flow term 
$\partial_x A \vec {u}$. The anharmonic chains to be studied have three conservation laws, hence $n=3$,
 and a matrix $A$  with  nondegenerate eigenvalues.
There are early studies of multi-component KPZ equations  \cite{EK93,DBBR01}, assuming however  $A=0$, which then leads to a scenario
very different from the one discussed here.

For the remainder of these lectures we will discuss one-dimensional  mechanical systems governed by Newton's equations of motion,
either point particles or discrete nonlinear wave equations. Usually they have three conserved fields. But  an example with $n=2$ will also be considered. The methods to be developed can also be used  for multi-component ASEP or other stochastic models with several conservation laws, see \cite{StSp14,PSSS15}.
 We will spend some time to define the models, to argue how the coupled system 
\eqref{0.1} of conservation laws arises, and to explain how the model dependent coefficients are computed. A separate task will be to extract out of  a system of nonlinear stochastic conservation laws concrete predictions which may be checked through molecular dynamics  simulations.

 \section{One-dimensional fluids and anharmonic chains}
 \label{sec7}
As microscopic model we focus on a classical fluid on the line consisting of particles with positions $q_j$ and momenta $p_j$, $j = 1,...,N$, $q_j,p_j \in \mathbb{R}$,
possible boundary conditions to be delayed momentarily. We use units such that the mass of the particles equals 1. Then the Hamiltonian is of the standard form,
\begin{equation}\label{3.1}
H_N^\mathrm{fl} = \sum_{j=1}^N \tfrac{1}{2} p_j^2 + \tfrac{1}{2}\sum_{i \neq j = 1}^NV(q_i - q_j)\,, 
\end{equation}
with pair potential $V(x) = V(-x)$. The potential may have a hard core and otherwise is assumed to be short ranged.
The dynamics for long range potentials is of independent interest \cite{Ru14}, but not discussed here.  The three conserved fields are density, momentum, and energy. One might want to add a periodic external potential. Then momentum is no more conserved and the dynamical properties will change dramatically.

A substantial simplification is achieved by assuming a hard core of diameter $a$, \textit{i.e.}  $V(x) = \infty$ for $|x| < a$, and restricting the range of the smooth part of the potential to at most $2a$. Then the particles maintain their order,  $q_j \leq q_{j+1}$, and in addition only nearest neighbor particles interact. Hence $H_N^\mathrm{fl}$
simplifies to
\begin{equation}\label{3.2}
H_N = \sum_{j=1}^N  \tfrac{1}{2}p_j^2 + \sum_{j = 1}^{N-1}V(q_{j+1} - q_j)\,.
\end{equation}
As a, at first sight very different,  physical realization,  we could interpret $H_N$ as describing particles in one dimension coupled through anharmonic springs which is then usually referred to as anharmonic chain.

In the second interpretation the spring potential can be more general than anticipated so far. No ordering constraint is  required and the potential does not have to be even. To have
well defined thermodynamics the chain is pinned at both ends as $q_1 =0$ and $q_{N+1} = \ell N$. It is convenient to introduce the stretch $r_j = q_{j+1} - q_j$. Then the boundary condition corresponds to the microcanonical constraint
\begin{equation}\label{3.3}
\sum_{j=1}^{N} r_j = \ell N\,. 
\end{equation} 
Switching   to canonical equilibrium according to the standard rules, one the arrives at the  obvious condition  of 
a finite partition function
\begin{equation}\label{3.4}
Z(P,\beta) =  \int_\mathbb{R}dx \,\mathrm{e}^{-\beta(V(x) +Px)} < \infty \,, 
\end{equation} 
using the standard convention that the integral is over the entire real line. Here $\beta  > 0 $ is the inverse temperature and $P$ is the thermodynamically conjugate variable to the stretch.
By partial integration
\begin{equation}\label{3.5}
 P = -Z(P,\beta)^{-1} \int_\mathbb{R} dx V'(x)\,\mathrm{e}^{-\beta(V(x) +Px)}  \,, 
\end{equation} 
implying that $P$ is the average force in the spring between two adjacent particles, hence identified as thermodynamic  pressure. To have a finite partition function, a natural condition on the potential is to be bounded from below 
and to have a one-sided linear bound as 
$V(x) \geq a_0 + b_0|x|$ for either $x> 0$ or $x < 0$ and $b_0 >0$. Then there is a non-empty interval $I(\beta)$ such that $Z(P,\beta) < \infty $ for $P\in I(\beta)$. For the particular case of  a hard-core fluid one has to impose  $P >0$.
\smallskip\\
\textit{Note:} The sign of $P$ is chosen such that for a gas of hard-point particles one has the familiar ideal gas law $P = 1/\beta \ell$. The chain tension is $-P$. \smallskip

Famous examples are the harmonic chain, $V_\mathrm{ha}(x) = x^2$, the Fermi-Pasta-Ulam (FPU) chain,
$V_\mathrm{FPU}(x) = \tfrac{1}{2}x^2 + \tfrac{1}{3}\alpha x^3 + \tfrac{1}{4}\beta x^4$, in the historical notation
\cite{FPU56}, and the Toda chain \cite{To60}, $V(x) = \mathrm{e}^{-x}$, in which case $P>0$ is required. The harmonic chain, the Toda chain, and the hard-core potential, $V_\mathrm{hc}(x) = \infty$ for $|x| < a$ and $V_\mathrm{hc}(x) = 0$ for $|x| \geq a$, are in fact integrable systems which have a very different correlation structure and will not be discussed here. Except for the harmonic chain, one simple way to break integrability is to assume alternating masses, say $m_j = m_0$ for even $j$ and $m_j = m_1$ for odd $j$.

We will mostly deal with anharmonic chains described by the Hamiltonian \eqref{3.2}, including  one-dimensional hard-core fluids with a sufficiently small potential range. There are two good reasons.  
Firstly, amongst the large body of molecular dynamics simulations there is not a single one which deals
with an ``honest'' one-dimensional fluid. To be able to reach large system sizes all simulations are performed for anharmonic chains. 
Secondly, from a theoretical perspective, the equilibrium measures of anharmonic chains are particularly simple in being
of product form in momentum and stretch variables. Thus material parameters, as compressibility and sound speed, can be
expressed in terms of one-dimensional integrals involving the Boltzmann factor $\mathrm{e}^{-\beta(V(x) +Px)}$, $V(x)$,
and $x$.

The dynamics of the anharmonic chain is governed by 
\begin{equation}\label{3.6}
\frac{d}{dt} q_j=p_{j}\,,\qquad \frac{d}{dt}{p}_j=V'(q_{j+1} - q_j) - V'(q_j - q_{j-1})\,.
\end{equation}
This can be viewed as the discretization of the nonlinear wave equation
\begin{equation}\label{3.6a}
\partial_t^2u(x,t) =\partial_xV'(\partial_x u(x,t)) \,,
\end{equation}
which for the harmonic potential reduces to the linear wave equation. In this physical interpretation
$\{q_j, j =1,...,N\}$ is the discretized displacement field $u(x)$. Throughout we will stick to the lattice field theory point of view.
For the initial conditions we choose a lattice cell of length $N$ and require
\begin{equation}\label{3.7}
q_{j+N} = q_j +\ell N\,,\qquad p_{j+N} = p_j
\end{equation}
for all $j \in \mathbb{Z}$. This property is preserved under the dynamics and thus properly mimics a system of finite length $N$. The stretches are then $N$-periodic, $r_{j+N} = r_j$, and the single cell dynamics is given by 
 \begin{equation}\label{3.8}
\frac{d}{dt} r_j=p_{j+1}-p_j\,,\qquad \frac{d}{dt}{p}_j=V'(r_j)-V'(r_{j-1})\,,
\end{equation}
$ j=1,\ldots,N$, together with the periodic boundary conditions
 $p_{1+N} = p_1$, $r_ 0= r_N$ and the constraint \eqref{3.3}. 
Through the stretch there is a coupling to the right neighbor and through the momentum a coupling to the left neighbor.
The potential is defined only up to translations, since the dynamics does not change under a simultaneous shift of $V(x)$ to $V(x-a)$ and $r_j$ to $r_j+a$, in other words, the potential can be shifted by shifting the initial $r$-field.
Note that our periodic boundary conditions are not identical to  fluid particles moving on a ring, but they may
become so for large system size when length fluctuations become negligible.

Before proceeding let us be more specific on the link between fluids and anharmonic chains. To avoid the issue of boundary conditions we start from the infinitely extended system. First of all, for a potential as $V(x) = x^2 + x^4$ the lattice field theory point of view is the natural option. Similarly for $V(x) =1, |x| < a $, and $V(x) = 0, |x| \geq a$,  unlabeled particles moving on the real line is the obvious choice,
$q_j$ then being the physical position of the $j$-th particle. So let us consider a potential, for which both fluid and solid picture are meaningful.
Clearly, given  \eqref{3.6} also Eq. \eqref{3.8} is satisfied. In reverse order, given the solution to \eqref{3.8}, we still have to fix the value of $q_0$, say. Then $q_0(t)$ follows from  \eqref{3.6} and one thereby reconstructs all other positions. Thus up to the choice of 
$q_0$ the two dynamics are identical. However the observables for a hydrodynamic theory will be different.  In case of a fluid, momentum and energy are attached to a particle. For example, the momentum field is defined by
\begin{equation}\label{3.8a}
\mathsf{u}_\mathrm{fl}(x,t) = \sum_ {j}\delta (x - q_j(t))p_j(t),
\end{equation}
while for the lattice field theory it is merely $p_j(t)$. Both fields are locally conserved.  The dynamical 
correlator for the fluid, $\langle \mathsf{u}_\mathrm{fl}(x,t)\mathsf{u}_\mathrm{fl}(0,0)\rangle$, differs from the corresponding lattice correlator,  $\langle p_j(t)p_0(0)\rangle$, and there is no simple rule to transform one into the other. On the other hand the hydrodynamic theory to be developed could as well be carried through for fluids. Structurally both theories will be of the form
of Eq. \eqref{0.1}. By universality we thus expect to have the same behavior on large scales.

 We return to anharmonic chains and note that  equations \eqref{3.8}  are already of conservation type. Hence
\begin{equation}\label{3.9}
\frac{d}{dt}\sum^N_{j=1} r_j=0\,,\quad \frac{d}{dt}\sum^N_{j=1} p_j=0\,.
\end{equation}
We define the local energy by 
\begin{equation}\label{3.10}
e_j=\tfrac{1}{2}p^2_j + V(r_j)\,.
\end{equation}
Then its local conservation law reads
\begin{equation}\label{3.11}
\frac{d}{dt}e_j= p_{j+1} V'(r_j)-p_j V'(r_{j-1})\,,
\end{equation}
implying that
\begin{equation}\label{3.12}
\frac{d}{dt}\sum^N_{j=1} e_j=0\,.
\end{equation}

The microcanonical equilibrium state is defined by the Lebesgue measure  constrained to a particular value of the conserved fields as
\begin{equation}\label{3.13}
\sum^N_{j=1} r_j = \ell N \,,\quad \sum^N_{j=1} p_j=\mathsf{u}N\,,\quad \sum^N_{j=1} \big(\tfrac{1}{2}p^2_j+ V(r_j)\big) = \mathfrak{e} N
\end{equation}
with $\ell$ the stretch,  $\mathsf{u}$ the momentum, and $\mathfrak{e}$ the total energy per particle. In our context the equivalence of ensembles holds and computationally it is of advantage to switch to the canonical ensemble with respect to all three constraints. Then the dual variable for the stretch $\ell$ is the pressure $P$, for the momentum the average momentum,
again denoted by $\mathsf{u}$, and for the total energy $\mathfrak{e}$ the inverse temperature $\beta$. 
For the limit of infinite volume the  symmetric choice $j \in [-N,...,N]$ is more convenient. In the limit $N \to \infty$ either under the canonical equilibrium state, trivially, or under the microcanonical ensemble, by the equivalence of ensembles,  the collection $(r_j,p_j)_{j\in\mathbb{Z}}$ are independent random variables. Their single site probability density is given by
\begin{equation}\label{3.14}
Z(P,\beta)^{-1} \mathrm{e}^{-\beta(V(r_j)+Pr_j)} (2\pi/\beta)^{-1/2} \mathrm{e}^{-\frac{1}{2}\beta (p_j -\mathsf{u})^2}\,.
\end{equation}
Averages with respect to~(\ref{3.14}) are denoted by $\langle\cdot\rangle_{P,\beta,\mathsf{u}}$. The dependence on the 
average momentum can be removed by a Galilei transformation. Hence we mostly work with $\mathsf{u} = 0$, in which case we merely drop the index $\mathsf{u}$. We also introduce the internal energy, $\mathsf{e}$, through 
$\mathfrak{e} = \tfrac{1}{2}\mathsf{u}^2 +\mathsf{e}$, which agrees with the total energy at $\mathsf{u} = 0$. The canonical free energy, at $\mathsf{u} = 0$, is defined by
\begin{equation}\label{3.15}
G(P,\beta)=-\beta^{-1} \big(-\tfrac{1}{2}\log\beta + \log Z(P,\beta)\big)\,.
\end{equation}
Then
\begin{equation}\label{3.16}
\ell =\langle r_0\rangle_{P,\beta}\,,\quad \mathsf{e}=\partial_\beta\big(\beta G(P,\beta)\big) -P \ell=\frac{1}{2\beta}+\langle 
V(r_0)\rangle_{P,\beta}\,.
\end{equation}
The relation (\ref{3.16}) defines $(P,\beta) \mapsto (\ell(P,\beta),\mathsf{e}(P,\beta))$, thereby the inverse map 
$(\ell, \mathsf{e}) \mapsto (P(\ell,\mathsf{e}),$ $  \beta(\ell,\mathsf{e}))$, and thus accomplishes the switch between
the microcanonical thermodynamic variables $\ell, \mathsf{e}$ and the canonical thermodynamic variables $P, \beta$.

It is convenient to collect the conserved fields as the $3$-vector
$\vec{g} = (g_1,g_2,g_3)$, 
\begin{equation}\label{3.17}
\vec{g}(j,t) = \big(r_j(t),p_j(t),e_j(t)\big) \,,
\end{equation}
$\vec{g}(j,0) = \vec{g}(j)$. Then the conservation laws are combined as
\begin{equation}\label{3.18}
\frac{d}{dt}\vec{g}(j,t) + \vec{\mathcal{J}}(j+1,t)  - \vec{\mathcal{J}}(j,t)=0 
\end{equation}
with the local current functions 
\begin{equation}\label{3.19}
\vec{\mathcal{J}}(j) = \big( -p_j,-V'(r_{j-1}), - p_jV'(r_{j-1})\big)\,.
\end{equation}

Very roughly, our claim is that, for suitable random initial data, on the mesoscopic scale the conservation
law \eqref{3.18} can be well approximated by a noisy conservation law of the form \eqref{0.1} with $n =3$.
This leaves the physical set-up widely unspecified. To be more concrete, also to have a well-defined
control through molecular dynamics simulations and a quantity of physical importance, we consider time correlation
functions in thermal equilibrium of the conserved fields.  
 They are defined by
\begin{equation}\label{3.20}
S_{\alpha\alpha'}(j,t)=\langle g_{\alpha}(j,t) g_{\alpha'}(0,0)\rangle_{P,\beta} - \langle g_{\alpha}(0,0)\rangle_{P,\beta} \langle g_{\alpha'}(0,0)\rangle_{P,\beta}\,,
\end{equation}
$\alpha,\alpha'=1,2,3$. The infinite volume limit has been taken already and the average is with respect to thermal equilibrium at $\mathsf{u} = 0$. It is known that such a limit exists \cite{Olla12}. Also the decay in $j$ is exponentially fast, but with a correlation length
increasing in time. 
In this context our central claim states that $S_{\alpha\alpha'}(j,t)$ can be well approximated by the stationary covariance
of a Langevin equation of the same form as in Eq. \eqref{0.1} with coefficients which still have to be computed. 
\begin{figure}
\centering
\includegraphics[width=0.8\textwidth]{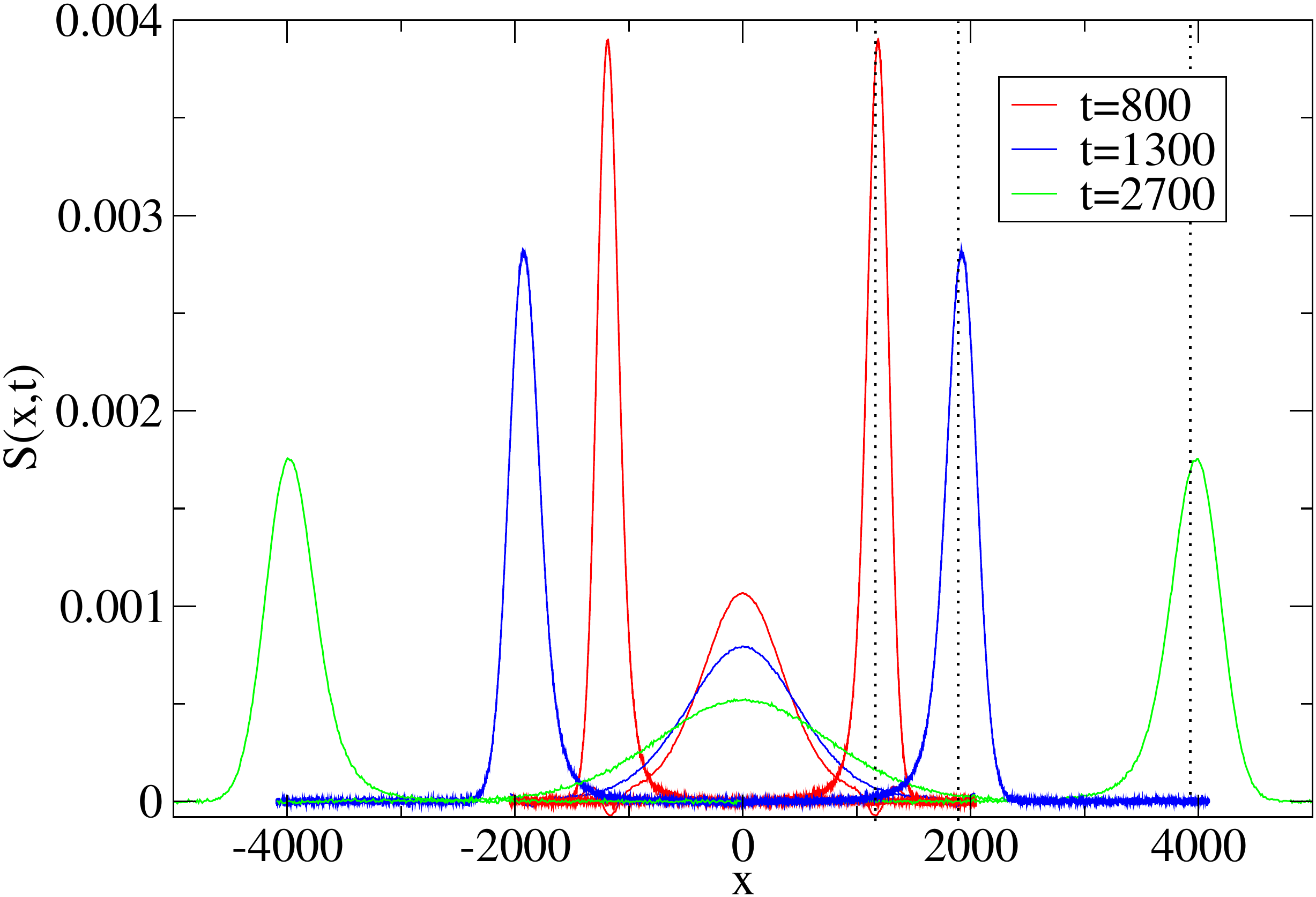}
\caption{Heat peak and two sound peaks, area normalized to 1, at times $t = 800,1300,2700$ for a FPU chain with $N=8192$, $V(x) = \tfrac{1}{2}x^2 + \tfrac{2}{3}x^3 + \tfrac{1}{4} x^4$, pressure $P= 1$, and inverse temperature $\beta = 2$. The sound speed is $c=1.45$. The heat peak has power law tails, which are cut off by the sound peaks.
According to the theory, asymptotically the sound peaks are symmetric relative $\sigma c t$ and have  stretched exponential tails as $\exp[-|x|^3]$.}
\label{fig3}
\end{figure}

Often it is convenient to regard $S(j,t)$, no indices, as a $3\times 3$ matrix. In general, $S(j,t)$ has certain symmetries, the first set resulting from space-time stationarity and the second set from time reversal, even for
$\alpha = 1,3$, odd for $\alpha = 2$, 
\begin{equation}\label{3.21}
S_{\alpha\alpha'}(j,t) = S_{\alpha'\alpha}(-j,-t)\,,\quad S_{\alpha\alpha'}(j,t) = (-1)^{\alpha+ \alpha'}S_{\alpha\alpha'}(j,-t) 
\,.
\end{equation}

At $t=0$ the average \eqref{3.20} reduces to a static average, which is easily computed.  In general the static susceptibility matrix is defined through
\begin{equation}\label{3.21a}
C = \sum_{j\in\mathbb{Z}} S(j,0)
\,.
\end{equation}
For anharmonic chains the fields are uncorrelated in $j$, hence
\begin{equation}\label{3.22}
S(j,0)=\delta_{j0} C
\end{equation}
with susceptibility matrix
\begin{equation}\label{3.23}
C=
\begin{pmatrix} \langle r_0;r_0\rangle_{P,\beta} & 0 & \langle r_0;V_0\rangle_{P,\beta} \\
                0 & \beta^{-1} & 0 \\
                \langle r_0;V_0\rangle_{P,\beta} & 0 & \frac{1}{2}\beta^{-2}+\langle V_0;V_0\rangle_{P,\beta}
\end{pmatrix}\,.
\end{equation}
Here, for $X$,$Y$ arbitrary random variables, $\langle X;Y\rangle=\langle XY\rangle-\langle X\rangle\langle Y\rangle$ denotes the second cumulant
and $V_0 = V(r_0)$, following the same notational convention as for $e_0$.
Note that the conservation law implies the zeroth moment sum rule
\begin{equation}\label{3.24}
\sum_{j \in \mathbb{Z}}S(j,t) = \sum_{j \in \mathbb{Z}}S(j,0) = C\,.
\end{equation}

Since our theoretical discussion will be somewhat lengthy, we indicate already now in which direction we are heading.
Fig. \ref{fig3} displays  a molecular dynamics simulation of a FPU chain with potential $V(x) = \tfrac{1}{2}x^2 + \tfrac{2}{3}x^3 + \tfrac{1}{4} x^4$ at inverse temperature $\beta = 2$ and pressure $P= 1$ \cite{DDSMS14}. The system size is $N = 8192$. Plotted is a time sequence of a generic matrix element of $j \mapsto S(j,t)$ with a normalization such that the area under each peak equals $1$. There are two sound peaks, one moving to the right and its mirror image to the left
with sound speed $c= 1.45$. In addition there is a peak standing still, which for thermodynamic reasons is called the heat peak. The peaks broaden in time as $t^{2/3}$ for sound and as $t^{3/5}$ for heat.
The average is over $10^7$ samples drawn from the canonical distribution \eqref{3.14}.

\section{Discrete nonlinear Schr\"{o}dinger equation}\label{sec8}
In parallell to anharmonic chains, it  is instructive to discuss as second microscopic model the nonlinear Schr\"{o}dinger equation on the one-dimensional lattice (DNLS). As a novel feature, the large scale dynamics at high temperatures is diffusive, while at low temperatures propagating modes appear because of  an additional almost conserved field.

For DNLS the lattice field is $\psi_j \in \mathbb{C}$ and $\psi$, $\psi^*$ are canonically conjugate fields, $^*$ denoting complex conjugation. The Hamiltonian reads
\begin{equation}\label{4.1}
H = \sum_{j=1}^N \big(\tfrac{1}{2}|\psi_{j+1} - \psi_j|^2 + \tfrac{1}{2}g |\psi_j|^4 \big)
\end{equation}
with periodic boundary conditions. $g > 0$ is the coupling constant, a \emph{defocusing} nonlinearity.  
The dynamics is defined through
\begin{equation}
\mathrm{i} \, \frac{d}{d t} \psi_j = \partial_{\psi_j^*} H
\end{equation}
and thus
\begin{equation}
\label{4.2}
\mathrm{i} \, \frac{d}{d t} \psi_j = -\tfrac{1}{2} \Delta \psi_{j} + g\,\lvert\psi_j\rvert^2\,\psi_j
\end{equation}
with the lattice Laplacian $\Delta = -\partial ^\mathrm{T}\partial$ and $\partial \psi_j = \psi_{j+1} - \psi_j$. 

The DNLS has two obvious locally conserved fields, density and energy, 
\begin{equation}\label{4.3}
\rho_j = \lvert\psi_j\rvert^2\,,\qquad e_j = \tfrac{1}{2} \lvert\psi_{j+1} - \psi_j\rvert^2 + \tfrac{1}{2}\,g\,\lvert\psi_j\rvert^4\,.
\end{equation}
According to the discussion in \cite{Ablowitz2004}, the DNLS is nonintegrable and one expects density and energy to be the only locally conserved fields. They satisfy the conservation laws
\begin{equation}\label{4.4}
\begin{split}
\frac{d}{d t} \rho_j(t) + \mathcal{J}_{\rho,j+1}(t) - \mathcal{J}_{\rho,j}(t) &= 0\,,\\[1ex]
\frac{d}{d t} e_j(t) + \mathcal{J}_{e,j+1}(t) - \mathcal{J}_{e,j}(t) &= 0\,,
\end{split}
\end{equation}
with density current
\begin{equation}\label{4.5}
\mathcal{J}_{\rho,j} = \tfrac{1}{2} \mathrm{i} \big( \psi_{j-1}\,\partial \psi_{j-1}^* - \psi_{j-1}^*\,\partial\psi_{j-1} \big)
\end{equation}
and energy current
\begin{equation}\label{4.6}
\mathcal{J}_{e,j} = \tfrac{1}{4} \mathrm{i} \big(\Delta\psi_{j}^* \,\partial\psi_{j-1} - \Delta\psi_{j}\,\partial\psi_{j-1}^* \big) + g \lvert\psi_j\rvert^2 \mathcal{J}_{\rho,j}\,.
\end{equation}

As a consequence the canonical equilibrium state is given by
\begin{equation}\label{4.7}
Z^{-1}\mathrm{e}^{-\beta(H - \mu \mathsf{N})}\prod_{j=1}^N d\psi_jd\psi_j^*\,,\quad \mathsf{N} =   \sum_{j=1}^N |\psi_j|^2\,,
\end{equation}
with  chemical potential $\mu \in \mathbb{R}$. Here we assume $\beta >0$. But  also negative temperature states, in the microcanonical ensemble, have been studied \cite{ILLP13,IPP13}. Then the dynamics is dominated by a coarsening process mediated through breathers.
In equilibrium, the $\psi$-field has high spikes at random locations embedded in a low noise background, which is very different from the positive temperature states considered here.

Canonically conjugate variables can also be introduced by splitting the wave function into its real and imaginary part as
\begin{equation}\label{4.8}
\psi_j = \tfrac{1}{\sqrt{2}} (q_j + \mathrm{i} p_j)\,.
\end{equation}
In these variables, the Hamiltonian reads
\begin{equation}
\label{4.9}
H = \sum_{j=1}^{N} \Big( \tfrac{1}{4 } \big( (\partial q_j)^2 + (\partial p_j)^2 \big) + \tfrac{1}{8}\,g \big(q_j^2+p_j^2\big)^2 \Big)\,.
\end{equation}\label{4.10}
The dynamics defined by Eq.~\eqref{4.2} is then identical to the hamiltonian system
\begin{equation}\label{4.11}
\frac{d}{d t} q_j = \partial_{p_j}H\,, \quad \frac{d}{d t} p_j = -\partial_{q_j} H\,.
\end{equation}
Note that $H$ is symmetric under the interchange $q_j \leftrightarrow p_j$.

It will be convenient to make a canonical (symplectic) change of variables to polar coordinates as
\begin{equation}\label{4.12}
\varphi_j = \mathrm{arctan}(p_j/q_j), \quad \rho_j = \tfrac{1}{2} \big(p_j^2+q_j^2\big)\,,
\end{equation}
which is equivalent to the representation
\begin{equation}\label{4.13}
\psi_j = \sqrt{\rho_j}\, \mathrm{e}^{\mathrm{i} \varphi_j}\,.
\end{equation}
In the new variables the phase space becomes $(\rho_j, \varphi_j) \in \mathbb{R}_+ \times S^1$, with $S^1$ the unit circle. The corresponding Hamiltonian is given by
\begin{align}\label{4.14}
H &= \sum_{j=1}^{N} \Big( \tfrac{1}{2 } \big( \sqrt{\rho_{j+1}\,\rho_j}\,2\,(1 - \cos(\varphi_{j+1} - \varphi_j)) + (\sqrt{\rho_{j+1}} - \sqrt{\rho_j})^2 \big) + \tfrac{1}{2}\,g\,\rho_j^2 \Big) \nonumber\\
&= \sum_{j=1}^{N} \big( - \sqrt{\rho_{j+1}\,\rho_j}\, \cos(\varphi_{j+1} - \varphi_j) + \rho_j + \tfrac{1}{2}\,g\,\rho_j^2 \big)\,.
\end{align}
The equations of motion read then
\begin{equation}\label{4.15}
\tfrac{d}{d t} \varphi_j = -\partial_{\rho_j} H\,, \quad \tfrac{d}{d t} \rho_j = \partial_{\varphi_j} H\,.
\end{equation}
From the continuity of $\psi_j(t)$ when moving through the origin, one concludes that at $\rho_j(t)= 0$ the phase jumps from $\varphi_j(t) $ to $ \varphi_j(t) + \pi$. The $\varphi_j$'s are angles and therefore position-like variables, while the $\rho_j$'s are actions and hence momentum-like variables. The Hamiltonian depends only on phase differences which implies the invariance under the global shift $\varphi_j \mapsto \varphi_j + \phi$.

In \eqref{4.1} the kinetic energy is chosen such that in the limit of zero lattice spacing one arrives at the continuum nonlinear Schr\"odinger equation on  $\mathbb{R}$. This is an integrable nonlinear wave equation, while the DNLS is nonintegrable. Thus lattice and continuum version show distinct dynamical behavior. 

 \section{Linearized Euler equations}
 \label{sec9}
Our goal is to predict the long time behavior of the correlations of the conserved fields, compare with \eqref{3.20}. 
$S_{\alpha\alpha'}(j,t)$ may be viewed as the response in the field $\alpha$ at $(j,t)$ to equilibrium perturbed in the field $\alpha'$ at $(0,0)$. One might hope to capture such a response on the basis of an evolution equation for the conserved fields when linearized at equilibrium. The most obvious macroscopic description are the Euler equations which are a fairly direct consequence of the conservation laws. One starts the system in a state of local equilibrium, which means to have the equilibrium parameters varying slowly on the scale of inter-particle distances. If the dynamics is sufficiently chaotic, such a situation
is expected to persist provided the parameters evolve according to the Euler equations. Their currents are thus obtained by averaging the
microscopic currents in a local equilibrium state. In other words, the Euler currents are defined through static expectations. More specifically, in the case of anharmonic chains we use the microscopic currents \eqref{3.19}. Then the average currents are
\begin{equation}\label{5.1}
\langle \vec{\mathcal{J}}(j)\rangle_{\ell,\mathsf{u},\mathfrak{e}} = \big(-\mathsf{u},P(\ell,\mathfrak{e}-\tfrac{1}{2}\mathsf{u}^2), \mathsf{u} P(\ell,\mathfrak{e}-\tfrac{1}{2}\mathsf{u}^2)\big) = \vec{\mathsf{j}}(\ell,\mathsf{u},\mathfrak{e})
\end{equation}
with $P(\ell,\mathsf{e})$ defined implicitly through~(\ref{3.16}). On the macroscopic scale the difference becomes $\partial_x$ and one arrives at 
the macroscopic Euler equations 
\begin{equation}\label{5.2}
\partial_t\ell -\partial_x \mathsf{u} =0\,,\quad
\partial_t \mathsf{u} +\partial_x P(\ell,\mathfrak{e}-\tfrac{1}{2}\mathsf{u}^2) =0\,,\quad
 \partial_t \mathfrak{e} +\partial_x \big( \mathsf{u} P(\ell,\mathfrak{e}-\tfrac{1}{2}\mathsf{u}^2)\big) =0
\end{equation}
with the three conserved fields depending on $x,t$. We refer to a forthcoming monograph \cite{Olla12}, where the validity of the Euler equations is proved up to the first shock. Since, as emphasized already, it is difficult to deal with deterministic chaos, the authors add random velocity exchanges between neighboring particles which ensure that the dynamics locally enforces the microcanonical state.   

For DNLS the situation is much simpler. The currents are symbolically of the from $\mathrm{i}(z - z^*)$ and the equilibrium state is invariant under complex conjugation. Hence the average currents vanish. We will later see that at low temperatures an additional almost conserved field emerges. Then the  Euler equations become nontrivial and have the same structure as in \eqref{5.2}.

We are interested here only in small deviations from equilibrium and therefore linearize the Euler equations  as $\ell+u_1(x)$, $0+u_2(x)$, $\mathsf{e}+u_3(x)$ to linear order in the deviations $\vec{u}(x)$. This leads to the  linear equation
\begin{equation}\label{5.3}
\partial_t \vec{u}(x,t)+A \partial_x \vec{u} (x,t)=0
\end{equation}
with
\begin{equation}\label{5.4}
A=
\begin{pmatrix} 0 & -1 & 0 \\
               \partial_\ell P & 0 & \partial_\mathsf{e}P \\
                0 & P & 0
\end{pmatrix}\,.
\end{equation}
Here, and in the following, the dependence of $A$, $C$ and similar quantities on the background values $\ell,\mathsf{u} =0,\mathsf{e}$, hence on $P,\beta$,
 is suppressed from the notation.
Beyond~(\ref{3.24}) there is the first moment sum rule which states that
\begin{equation}\label{5.5}
\sum_{j \in \mathbb{Z}}jS(j,t) = AC\,t\,.
\end{equation}
A proof, which in essence uses only the conservation laws and space-time stationarity of the correlations,
is given in \cite{Sp14}, see also see  \cite{To,Sc}. Microscopic properties enter only minimally. However, since $C = C^\mathrm{T}$ and $S(j,t)^\mathrm{T}
= S(-j,-t)$, Eq. \eqref{5.5} implies
the important relation
\begin{equation}\label{5.6}
  AC=(AC)^{\mathrm{T}}= CA^{\mathrm{T}}\,,
\end{equation}
with $^\mathrm{T}$ denoting transpose. Of course, \eqref{5.6} can be checked also directly from the definitions.
Since  $C>0$,  $A$ is guaranteed to have real eigenvalues
and a nondegenerate system of right and left eigenvectors. For $A$ one obtains the three eigenvalues $0, \pm c$
with 
\begin{equation}\label{5.7}
c^2= -\partial_{\ell} P+P \partial_\mathsf{e}P >0\,.
\end{equation}
Thus the solution to the linearized equation has three modes, one standing still, one right moving with velocity $c$ 
and one left moving with velocity $-c$. Hence we have identified the adiabatic sound speed as being equal to $c$.
 
\eqref{5.3} is a deterministic equation. But the initial data are random such that within our approximation
\begin{equation}\label{5.8}
\langle u_\alpha(x,0)u_{\alpha'}(x',0)\rangle = C_{\alpha\alpha'}\delta(x-x')\,.
\end{equation}
To determine the correlator $S(x,t)$ with such initial conditions is most easily achieved by introducing the linear transformation $R$ satisfying
\begin{equation}\label{5.9}
RAR^{-1} = \mathrm{diag}(-c,0,c)\,,\quad RCR^\mathrm{T} = 1\,.
\end{equation}
Up to trivial phase factors, $R$ is uniquely determined by these conditions. Explicit formulas are found in \cite{Sp14}.
Setting $\vec{\phi} = A \vec{u}$, one concludes
\begin{equation}\label{5.10}
\partial \phi_\alpha +c_\alpha \partial_x \phi_\alpha =0 \,,\quad \alpha= -1,0,1\,,
\end{equation}
with $\vec{c} = (-c,0,c)$. By construction, the random initial data have the correlator
\begin{equation}\label{5.11}
\langle \phi_\alpha(x,0)\phi_{\alpha'}(x',0)\rangle = \delta_{\alpha\alpha'} \delta(x-x')\,.
\end{equation}
Hence
\begin{equation}\label{5.12}
\langle \phi_\alpha(x,t)\phi_{\alpha'}(0,0)\rangle = \delta_{\alpha\alpha'}\delta(x - c_\alpha t)\,.
\end{equation}
We transform back to the physical fields. Then in the continuum approximation, at the linearized level,
\begin{equation}\label{5.13}
S(x,t) =R^{-1}\mathrm{diag}\big(\delta(x + ct),\delta(x ),\delta(x - c t)\big)R^{-\mathrm{T}}
\end{equation}
with $R^{-\mathrm{T}} = (R^{-1})^\mathrm{T}$.

Rather easily we have gained a crucial insight: $S(j,t)$ has three peaks which separate linearly in time.
For example, $S_{11}(j,t)$ has three sharp peaks moving with velocities $\pm c,0$.  
The peak standing still, velocity $0$, is called the heat peak and the two peaks moving with velocity $\pm c$ are called the sound peaks. 
Physically and as supported by the molecular dynamics simulation displayed in Fig. \ref{fig3}, one expects such peaks not to be strictly sharp, but to broaden in the course of time because of dissipation.  This issue will have to be explored in great detail. It follows from the zeroth moment sum rule that the area under each peak is preserved in time
and thus determined through \eqref{5.13}. Hence the   weights can be computed from the
matrix $R^{-1}$, usually called Landau-Plazcek ratios.  A Landau-Placzek ratio could  vanish,
either accidentally or by a particular symmetry.  An example is the momentum correlation $S_{22}(j,t)$. Since $ (R^{-1})_{20}=0$
always, its central peak is absent.

For integrable chains each conservation law generates a peak. Thus, \textit{e.g.}, $S_{11}(j,t)$ of the Toda chain is expected to have a broad spectrum expanding ballistically, rather than consisting of three sharp peaks.

\section{Linear fluctuating hydrodynamics}\label{sec10}
The broadening of the peaks results from random fluctuations in the currents, which tend to be uncorrelated in space-time.  Therefore the crudest model would be to assume that the current statistics  is space-time Gaussian white noise. In principle, the noise components could be correlated. But since the stretch current is itself conserved, its fluctuations will be taken care of by the momentum equation.
Momentum and energy currents have different signature under time reversal, hence their cross correlation vanishes.
As a result, there is a fluctuating momentum current of strength $\sigma_\mathsf{u}$ and an independent energy current 
of strength $\sigma_\mathsf{e}$. According to Onsager, noise is linked to dissipation as modeled by a diffusive term. Thus the linearized equations \eqref{5.3} are extended to 
\begin{equation}\label{6.1}
\partial_t \vec{\mathsf{u}}(x,t)+\partial_x \big( A \vec{\mathsf{u}} (x,t) - \partial_x  D\vec{\mathsf{u}}(x,t) + B\vec{\xi} (x,t)\big)=0\,.
\end{equation} 
Here $\vec{\xi} (x,t)$ is standard white noise with covariance
\begin{equation}\label{6.2}
\langle\xi_\alpha(x,t) \xi_{\alpha'}(x',t')\rangle= \delta_{\alpha\alpha'} \delta(x-x') \delta(t-t')
\end{equation}
and, as argued,  the noise strength matrix is diagonal as 
\begin{equation}\label{6.3}
B = \mathrm{diag}(0, \sigma_\mathsf{u},\sigma_\mathsf{e})\,.
\end{equation}
To distinguish the linearized Euler equations \eqref{5.3} from the Langevin equations \eqref{6.1}, we use 
$\vec{\mathsf{u}} = (\mathsf{u}_1,\mathsf{u}_2,\mathsf{u}_3)$ for the fluctuating fields.

The stationary measures for \eqref{6.1} are
spatial white noise with arbitrary mean. Since  small deviations from uniformity are considered, we always impose mean zero.  
Then  the components are correlated as
\begin{equation}\label{6.4}
\langle\mathsf{u}_\alpha(x) \mathsf{u}_{\alpha'}(x')\rangle= C_{\alpha\alpha'} \delta(x-x')\,.
\end{equation}
Stationarity relates the linear drift and the noise strength through the steady state covariance as
\begin{equation}\label{6.5}
- (AC -CA^\mathrm{T})\partial _x + (DC + CD^\mathrm{T})\partial_x^2 = BB^\mathrm{T} \partial_x^2\,.
\end{equation}
The first term vanishes by \eqref{5.6} and the diffusion matrix is uniquely determined as
\begin{equation}\label{6.6}
D=
\begin{pmatrix} 0 & 0 & 0 \\
              0 & D_\mathsf{u} & 0 \\
                
\tilde{D}_\mathsf{e}  & 0 & D_\mathsf{e}
\end{pmatrix}
\end{equation}
with $\tilde{D}_\mathsf{e} = - \langle r_0;V_0\rangle_{P,\beta} \langle r_0;r_0\rangle_{P,\beta}^{-1}D_\mathsf{e}$. Here $D_\mathsf{u} >0$ is the momentum and $D_\mathsf{e}> 0 $ the energy diffusion coefficient, which are related to the noise strength as
\begin{equation}\label{6.7}
 \sigma_\mathsf{u}^2 = \langle p_0;p_0\rangle_{P,\beta} D_\mathsf{u} \,,\quad \sigma_\mathsf{e}^2 =   \langle e_0;e_0\rangle_{P,\beta}
 D_\mathsf{e}\,.
\end{equation}

As an insert we return to the DNLS. To distinguish from $A,B,C,D$ for an anharmonic chain, we use $\mbox{\textsl{A,\,B,\,C,\,D}}$ as $2 \times 2$ matrices. There   are two conserved fields, but the Euler currents vanish. Hence linear fluctuating hydrodynamics takes the form
\begin{equation}\label{6.8}
\partial_t \vec{\mathsf{u}}(x,t)+\partial_x\big( - \partial_x  \mbox{\textsl{D}}\vec{\mathsf{u}}(x,t) + \mbox{\textsl{B}}\vec{\xi} (x,t)\big)=0
\end{equation}   
with two components, index 1 for density and index 2 for energy. The matrices $\mbox{\textsl{B,C}}$ satisfy the fluctuation relation
$\mbox{\textsl{DC}} +\mbox{\textsl{CD}} = \mbox{\textsl{BB}}^\mathrm{T}$, the susceptibility matrix $\mbox{\textsl{C}}$ being defined as in \eqref{3.21a}. Since the Hamiltonian in
\eqref{4.14} contains nearest neighbor couplings, this matrix is not as readily computed as in the case of anharmonic chains.
Physically there is a  more important difference. One expects that $\mbox{\textsl{DC}}$ can be written in terms of an time-integral over the total current-current correlation matrix. While not computable in analytic form, $\mbox{\textsl{D}}$ is thus a uniquely defined matrix.
On the other hand in \eqref{6.1} the diffusion matrix $D$ is a phenomenologically introduced coefficient. As will be argued, the
true long time behavior will depend only on $C$, which is uniquely defined, and not on $B,D$ separately, thus circumventing the arbitrariness in $D$.
\begin{figure}[!ht]
\centering
\subfloat[density $S_{\rho \rho}(j,t)$]{
\includegraphics[width=0.3\textwidth]{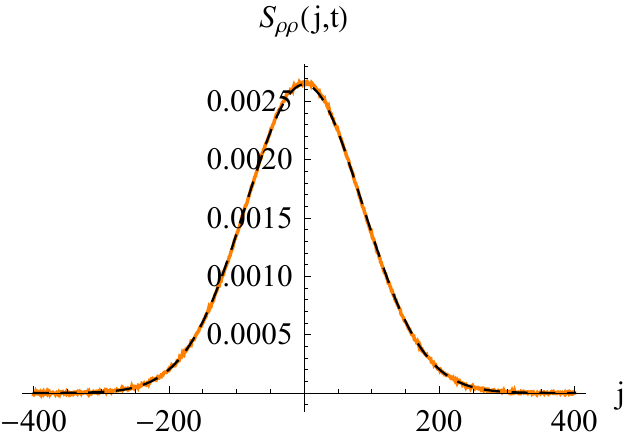}}
\hspace{0.02\textwidth}
\subfloat[density-energy $S_{\rho e}(j,t)$]{
\includegraphics[width=0.3\textwidth]{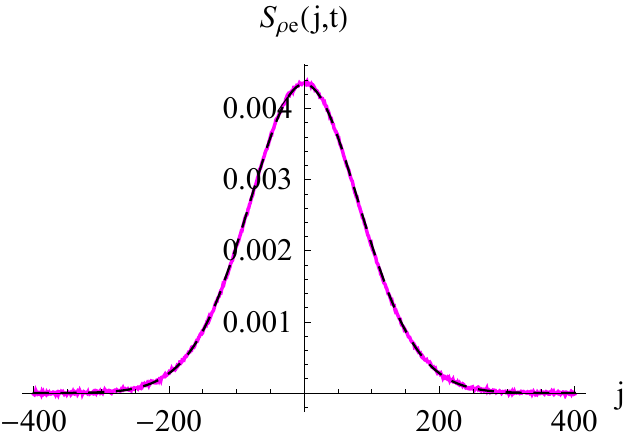}}
\hspace{0.02\textwidth}
\subfloat[energy $S_{e e}(j,t)$]{
\includegraphics[width=0.3\textwidth]{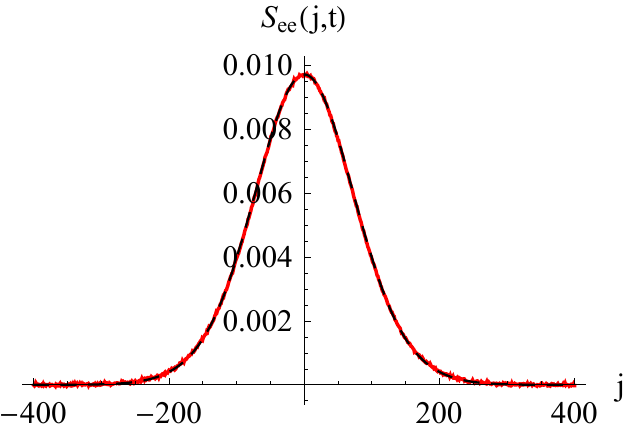}}
\caption{Equilibrium time-correlations $S_{\alpha \alpha'}(j,t)$ of density and energy  for the discrete NLS with $g=1$. The initial states are drawn from the canonical ensemble \eqref{4.7} with inverse temperature $\beta = 1$
and $\langle \rho_j\rangle = 1$. The black dashed curves are the entries on the right of Eq.~\eqref{6.9a}.}
\label{fig4}
\end{figure}

It is not difficult to solve \eqref{6.8}. We note that $\mbox{\textsl{D}}$ is in general not symmetric, but it has strictly positive eigenvalues.
This leads to the stationary two-point function
\begin{equation}\label{6.9}
\langle u_{\alpha}(x,t) u_{\alpha'}(0,0)\rangle = S_{\alpha\alpha'}(x,t) = \int_{\mathbb{R}} d k\, \mathrm{e}^{-\mathrm{i}2\pi k x} \big(\mathrm{e}^{-{(2\pi k)^2 }\mbox{\scriptsize{\textsl{D}}}\,t} \mbox{\textsl{C}} \big)_{\alpha\alpha'}\,,
\end{equation}
$\alpha,\alpha' = 1,2$. 
The corresponding lattice correlator reads
\begin{equation}\label{6.10a}
S(j, t) =
\begin{pmatrix}
\langle \rho_j(t) ; \rho_0(0)\rangle& \langle \rho_j(t); e_0(0) \rangle \\
\langle  e_j(t); \rho_0(0)\rangle & \langle e_j(t) ; e_0(0)\rangle\\
\end{pmatrix} \,.
\end{equation}
Working out the Fourier transform, one arrives at the prediction
\begin{equation}\label{6.9a}
S(j, t) \simeq \tfrac{1}{\sqrt{{4 \pi} \mbox{\scriptsize{\textsl{D}}}\, t}} \, \mathrm{e}^{-j^2/({4}\!\! \mbox{
\scriptsize{\textsl{D}}}\,t)} \,\mbox{\textsl{C}}\,.
\end{equation}

\begin{figure}[!ht]
\centering
\includegraphics[width=0.8\textwidth]{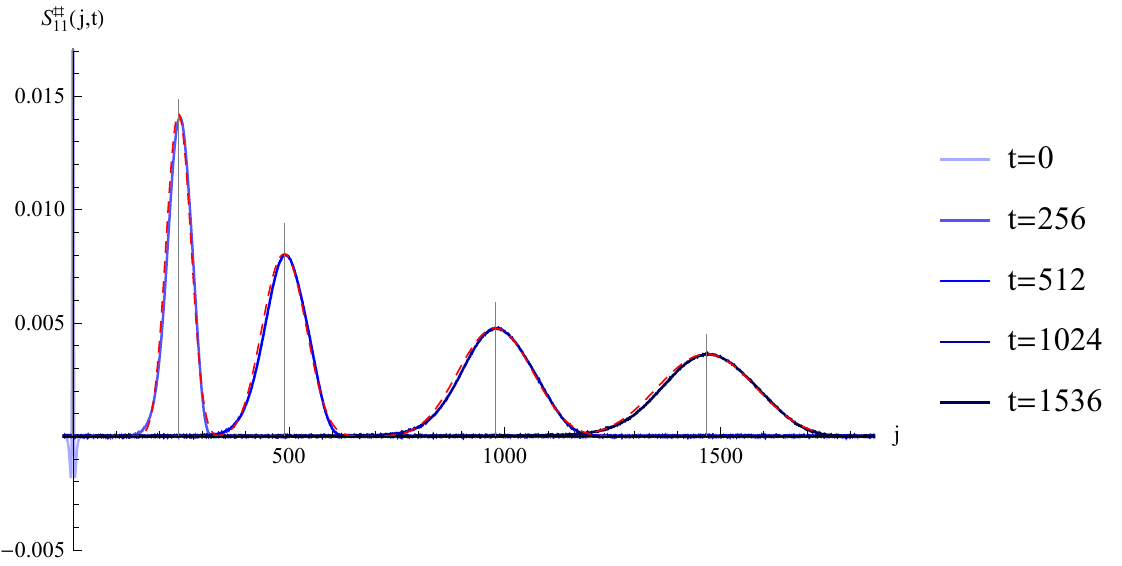}
\caption{Time sequence of the right moving sound peak, $S^{\sharp}_{11}(j,t)$, of the discrete NLS with 
parameters $N = 4096$, $g = 1$, $\beta = 15$, and $\langle \rho_j\rangle = 1$ corresponding to $\mu = 1.025$. The gray vertical lines show the theoretically predicted sound speed $c$ and the dashed red lines are the stationary KPZ scaling functions.}
\label{fig5}
\end{figure}

Such a prediction can be checked numerically, for which purpose we performed a molecular dynamics simulation  with system size system $N = 4096$ and parameters $g = 1$, $\beta = 1$, and $\langle \rho_j \rangle = 1$
\cite{MeSp15}. The susceptibility and diffusion matrix are obtained as 
\begin{equation}\label{6.10}
\mbox{\textsl{C}} =
\begin{pmatrix}
0.580 & 0.907 \\
0.907 & 1.848 \\
\end{pmatrix} \,,\qquad
\mbox{\textsl{D}}= \begin{pmatrix}
3.079 & -0.350 \\
2.298 &  0.897 \\
\end{pmatrix}\,.
\end{equation}
Both density and energy are even under time reversal, which explains that cross correlations are permitted.
The Gaussian fit is essentially perfect, compare with Fig. \ref{fig4}. $\mbox{\textsl{D}}$ is measured at the longest available time $t =1536$. However,
for $\beta = 15$ the behavior is drastically different, as confirmed by Fig. \ref{fig5}, which shows a right-moving sound peak broadening as $t^{2/3}$. The heat peak has in comparison a much smaller amplitude and is not resolved.
For a theoretical explanation, we will have to wait until Sec. \ref{sec13}.

We return to anharmonic  chains. Based on \eqref{6.1} one computes the stationary space-time covariance, which most easily is written in Fourier space,
\begin{equation}\label{6.11}
S_{\alpha\alpha'}(x,t) = \langle \mathsf{u}_\alpha(x,t)  \mathsf{u}_{\alpha'}(0,0)\rangle  = \int_\mathbb{R} dk\,\mathrm{e}^{\mathrm{i} 2 \pi
kx} \big(\mathrm{e}^{-\mathrm{i}t 2\pi kA - |t|(2\pi k)^2 D}C\big)_{\alpha\alpha'}\,. 
\end{equation}
To extract the long time behavior it is convenient to transform to normal modes. But before, we have to introduce
a more systematic notation. We will use throughout the superscript $^\sharp$ for a normal mode quantity. Thus for the anharmonic chain
\begin{equation}\label{6.13}
S^\sharp(j,t) = RS(j,t)R^\mathrm{T}\,,\quad S_{\alpha\alpha'}^\sharp(j,t)  = \langle (R\vec{g})_\alpha(j,t) ;(R\vec{g})_{\alpha'}(0,0)\rangle_{P,\beta}\,.
\end{equation}
The hydrodynamic fluctuation fields are defined on the continuum, thus functions of $x,t$, and we write
\begin{equation}\label{6.14}
 S_{\alpha\alpha'}(x,t)  = \langle \mathsf{u}_\alpha(x,t)  \mathsf{u}_{\alpha'}(0,0)\rangle\,,\quad S^\sharp(x,t) = RS(x,t)R^\mathrm{T}\,.
\end{equation}
Correspondingly $A^\sharp = R A R^{-1} = \mathrm{diag}(-c,0,c)$, $D^\sharp= R D R^{-1}$, $B^\sharp = RB$.
Note that $\vec{\mathsf{u}}(x,t)$ will change its meaning when switching from linear to nonlinear fluctuating hydrodynamics.

In normal mode representation Eq. \eqref{6.14} becomes
\begin{equation}\label{6.15}
S^\sharp(x,t)   = \int_\mathbb{R} dk\,\mathrm{e}^{\mathrm{i} 2 \pi
kx} \mathrm{e}^{-\mathrm{i}t 2\pi k A^\sharp- |t|(2\pi k)^2 D^\sharp}\,.
\end{equation}
The leading term, $\mathrm{i}t 2\pi kA^\sharp$, is diagonal, while the diffusion matrix $D^\sharp$ couples the components. 
But for large $t$ the peaks are far apart and the cross terms become small. More formally we split
$D^\sharp = D_\mathrm{dia} + D_\mathrm{off}$ and regard the off-diagonal part $D_\mathrm{off}$
as perturbation. When expanding, one notes that the off-diagonal terms carry an oscillating factor with frequency
$c_\alpha - c_\mathrm{\alpha'}$, $\alpha \neq \alpha'$. Hence these terms decay quickly and
 \begin{equation}\label{6.16}
S^\sharp_{\alpha\alpha'}(x,t)   \simeq \delta_{\alpha\alpha'} \int_\mathbb{R} dk\,\mathrm{e}^{\mathrm{i} 2 \pi
kx} \mathrm{e}^{-\mathrm{i}t 2\pi k c_\alpha- |t|(2\pi k)^2 D^\sharp_{\alpha\alpha}}
\end{equation}
for large $t$. Each peak has a Gaussian shape function which broadens as $(D^\sharp_{\alpha\alpha}|t|)^{1/2}$.

Besides the peak structure, we have gained a second important insight. Since the peaks travel with distinct velocities,
on the linearized level the three-component system decouples into three scalar equations,
provided it is written in normal modes. The two sound peaks are mirror images of each other and broaden with 
$D^\sharp_{11}= D^\sharp_{-1-1}$ and the heat peak broadens with $D^\sharp_{00}$.

\section{Second order expansion, \\nonlinear fluctuating hydrodynamics}\label{sec11}
From the experience with the scalar case, we know that at least the second order expansion of the Euler currents has to
be included. In this case the stationary measure is spatial white noise and the quadratic nonlinearity yields the dynamic exponent $z=3/2$.
Thus, we retain dissipation and noise in \eqref{6.1},  but expand the Euler currents of \eqref{5.1} beyond first order to
now include second order in $\vec{u}$, which turns \eqref{6.1} into the equations of nonlinear fluctuating hydrodynamics, \begin{eqnarray}\label{7.1}
&&\hspace{-20pt}\partial_t \mathsf{u}_1 -\partial_x \mathsf{u}_2 =0\,, \nonumber \\
&&\hspace{-20pt}\partial _t \mathsf{u}_2 +\partial_x\big((\partial_\ell P) \mathsf{u}_1 +(\partial_\mathsf{e}P) \mathsf{u}_3 + \tfrac{1}{2}
(\partial_\ell^2 P) \mathsf{u}_1^2 -\tfrac{1}{2} (\partial_\mathsf{e}P)\mathsf{u}_2^2 +\tfrac{1}{2}(\partial^2_\mathsf{e}P) \mathsf{u}_3^2 +
(\partial_\ell\partial_\mathsf{e}P) \mathsf{u}_1\mathsf{u}_3\nonumber\\
&&\hspace{260pt} - D_\mathsf{u}\partial_x \mathsf{u}_2 + \sigma_\mathsf{u}\xi_2\big) = 0\,,\nonumber\\
&&\hspace{-20pt}\partial _t \mathsf{u}_3 +\partial_x\big(P\mathsf{u}_2 +(\partial_\ell P) \mathsf{u}_1\mathsf{u}_2 +(\partial_\mathsf{e} P)\mathsf{u}_2\mathsf{u}_3 - \tilde{D}_\mathsf{e}
\partial_x \mathsf{u}_1 -D_\mathsf{e}\partial_x \mathsf{u}_3 +\sigma_\mathsf{e}\xi_3\big)=0\,.
\end{eqnarray} 
To explore their consequences is a more demanding task than solving the linear Langevin equation
and the results of the analysis will be more fragmentary.

To  proceed further, it is convenient to write \eqref{7.1} in vector form,
 \begin{equation}\label{7.2}
\partial_t \vec{\mathsf{u}}(x,t) +\partial_x \big(A \vec{\mathsf{u}}(x,t) +\tfrac{1}{2}\langle \vec{\mathsf{u}}, \vec{H} \vec{\mathsf{u}}\rangle -\partial_x D \vec{\mathsf{u}}(x,t) + B\vec{\xi}(x,t)\,\big)=0\,,
\end{equation}
where $\vec{H}$ is the vector consisting of the Hessians of the  currents with derivatives evaluated at the background values $(\ell,0,\mathsf{e})$,
\begin{equation}\label{7.3}
  H^\alpha_{\gamma\gamma'} =\partial_{\mathsf{u}_\gamma} \partial_{\mathsf{u}_{\gamma'}} \mathsf{j}_\alpha\,,\qquad
 \langle \vec{\mathsf{u}}, \vec{H} \vec{\mathsf{u}}\rangle = \sum^3_{\gamma,\gamma'=1}\vec{H}_{\gamma\gamma'} \mathsf{u}_\gamma \mathsf{u}_{\gamma'}\,.
\end{equation}
As for the linear Langevin equation we transform to normal modes through 
 \begin{equation}\label{7.4}
 \vec{\phi} = R \vec{\mathsf{u}}\,.
\end{equation}
 Then 
 \begin{equation}\label{7.5}
\partial_t \phi_\alpha + \partial_x \big(c_\alpha \phi_\alpha + \langle\vec{\phi}, G^{\alpha}\vec{\phi}\rangle-\partial_x(D^\sharp\vec{\phi})_\alpha+(B^\sharp\vec{\xi})_\alpha\big)=0
\end{equation}
with $D^\sharp = R D R^{-1}$, $B^\sharp = RB$. By construction $B^\sharp B^{\sharp\mathrm{T}} = 2 D^\sharp $. The nonlinear coupling constants, denoted by $\vec{G}$,
are defined by
\begin{equation}\label{7.6}
 G^\alpha = \tfrac{1}{2}\sum^3_{\alpha'=1}  R_{\alpha\alpha'}  R^{-\mathrm{T}} H^{\alpha'}R^{-1}
\end{equation}
with the notation $R^{-\mathrm{T}} = (R^{-1})^\mathrm{T}$.

Since derived from a chain, the couplings are not completely arbitrary, but satisfy the symmetries 
\begin{eqnarray}\label{7.7}
&&\hspace{-20pt}G^\alpha_{\beta\gamma} = G^\alpha_{\gamma\beta}\,, \quad G^\sigma_{\alpha\beta} 
= - G^{-\sigma}_{-\alpha-\beta}\,,\quad
 G^\sigma_{-10}  = G^\sigma_{01}  \,,  \nonumber\\
 &&\hspace{-20pt}G^0_{\sigma\sigma}  = - G^0_{-\sigma-\sigma}\,,\quad  
G^0_{\alpha\beta} = 0\,\,\,\, \mathrm{otherwise}. 
\end{eqnarray}
In particular note that  
\begin{equation}\label{7.8}
 G_{00}^0 = 0
 \end{equation} 
 always, while  $G^1_{11} = - G^{-1}_{-1-1}$ are generically different from 0. This property signals that the heat peak will behave differently from the sound peaks. The $\vec{G}$-couplings are listed in \cite{Sp14} and as a function of $P,\beta$ expressed in cumulants up to third order in $r_0,V_0$. The algebra is somewhat messy. But there is a short MATHEMATICA program available \cite{MeTUM} which, for given $P,\beta, V$, computes all coupling constants $G$, including the matrices
 $C,A,R$.
 
\section{Coupling coefficients, dynamical phase diagram}\label{sec12}
The coupling coefficients, $\vec{G}$, determine the long time behavior of the correlations of the conserved fields. 
A more  quantitative treatment will follow, but I first want to discuss some general features independent of the 
specific underlying microscopic model. For a given class of models, one can change the model parameters, which are either thermodynamic, like pressure, or mechanical, like a coefficient in the potential. For the purpose of the discussion,
I call both model parameters. The coupling coefficients are complicated functions of the model parameters.
So dynamical phase diagrams means the dynamical properties in dependence on model parameters as mediated through the $\vec{G}$-couplings.

We emphasize that the $\vec{G}$-couplings cannot distinguish between the microscopic dynamics being integrable or not. 
For example, the $\vec{G}$-couplings of the Toda lattice are not much different from nearby non-integrable chains. The assumption of chaotic microscopic dynamics is used already when writing down the Euler equations.

Our discussion assumes that the peak velocities are distinct, $c_\alpha \neq c_{\alpha'}$ for   $\alpha \neq \alpha'$. The degenerate case still remains to be  studied. One has to distinguish three types of couplings,\smallskip\\
-- self-couplings: $G^\alpha_{\alpha\alpha}$,\smallskip\\\
-- diagonal couplings, but not self:  $G^\alpha_{\alpha'\alpha'}$, $\alpha \neq \alpha'$,\smallskip\\\
-- off-diagonal couplings: $G^\alpha_{\alpha'\alpha''}$, $\alpha' \neq \alpha''$.\smallskip\\\
The off-diagonal couplings are irrelevant, in the sense that they do not show in the long time behavior. Of course, they may strongly modify the short term behavior. The rough argument is that for long times the peaks $\alpha'$ and $\alpha''$ have spatially a very small overlap and therefore  feed back little to the mode $\alpha$. On the other hand, if $\alpha' = \alpha''$, then the two peaks are on top of each other, and the peak $\alpha$ can still interact with them through having a slow spatial decay.

The dynamical phase diagram is characterized by the vanishing of some relevant couplings $G^\alpha_{\alpha'\alpha'}$. For $n=3$, there are $9$ relevant couplings and hence $2^9$ dynamical phases, at least in principle. Many of them will have coinciding properties and do not need to be distinguished. Within a given class of models generically not all phases are actually realized. In fact, anharmonic chains have only three dynamical phases and the DNLS has only one, even at low temperatures. It may happen, within a given class of models, that 
 $G^\alpha_{\alpha'\alpha'}$ has either a definite sign or vanishes identically. Thereby the richness of the phase diagram is strongly reduced. From a theoretical perspective, the most interesting case is a coupling $G^\alpha_{\alpha'\alpha'}$
with no definite sign. Then the model parameters can be changed so to make $G^\alpha_{\alpha'\alpha'}=0$. This could be so because of an additional symmetry. But it could simply be  an accidental zero for a function on a high-dimensional parameter space. 
Once such a zero is located, the theory predicts that at this zero the dynamical properties change dramatically, so to speak out of the blue, because the parameters as such do not look very different from neighboring parameter values.
Such a qualitative change is often more easily detectable, in comparison to a precise measurement of dynamical scaling exponents.

Anharmonic chains have special symmetries and not all possible couplings $\vec{G}$ can be realized. Listing the relevant parameters one has only four distinct parameters, $a_1,...,a_4$,
\begin{equation}\label{8.1}
 \mathrm{diag}(G^{-1}) = (-a_3,-a_2,-a_1)\,,\quad
 \mathrm{diag}(G^{0}) = (-a_4,0, a_4)\,,\quad \mathrm{diag}(G^{1})= (a_1,a_2,a_3)\,.
\end{equation}
From the explicit expression of $a_4$ one concludes that $a_4 >0$ and only three parameters remain. Their variation leads to three distinct phases,\medskip\\
(Phase 1) $G^1_{11} = - G^{-1}_{-1-1} \neq 0$. This is the generic phase. According to mode-coupling theory, to be discussed below, the two sound peaks broaden as KPZ, while the heat peak scales as symmetric  
L\'{e}vy $\tfrac{5}{3}$. \medskip\\
(Phase 2) $G^1_{11} =0$ and $a_1 =0, a_2 = 0$. Then the sound peaks decouple and broaden diffusively, while 
the heat peak scales as symmetric  
L\'{e}vy $\tfrac{3}{2}$. \medskip\\
(Phase 3) $G^1_{11} =0$ and either $a_1 \neq 0, a_2 \neq 0$  or $a_1 \neq 0, a_2 = 0$ or  $a_1 = 0, a_2 \neq 0$.
Since $G^0_{11} = - G^0_{-1-1} > 0$, all three peaks are cross-coupled. This leads to a heat peak which scales with 
symmetric  L\'{e}vy $\kappa$, a left sound peak which scales as maximally left asymmetric  L\'{e}vy $\kappa$,
and a right sound peak which scales as maximally right asymmetric  L\'{e}vy $\kappa$. Here $\kappa$ is the golden mean,
$\kappa = \tfrac{1}{2}( 1 + \sqrt{5})$.\medskip

To realize Phase 2 one
starts from the observation that 
the $\vec{G}$ coefficients are expressed through cumulants in $r_0$, $V_0$ . If the integrands are antisymmetric under reflection, many terms vanish.  The precise condition on the potential is to have some $a_0$,
$P_0$ such that
\begin{equation}\label{8.3}
V(x -a_0) +P_0x = V(-x-a_0) -P_0x\,
\end{equation}
for all $x$. Then for   $P=P_0$ and arbitrary $\beta$, one finds 
\begin{equation}\label{8.4}
G^1_{11} =0\,,\quad G^1_{-1-1} =  -G^{-1}_{11} = 0\,,\quad G^1_{00} =  -G^{-1}_{00} = 0\,,
\end{equation}
while $\sigma G^0_{\sigma\sigma} > 0$. The standard examples for \eqref{8.3} to hold are a FPU chain with no cubic interaction term, the $\beta$-chain, and the square well potential with alternating masses, both at zero pressure. 

Two examples of accidental zeros of  $G^1_{11}$ have been found recently for  the asymmetric FPU potential $V(x)
= \tfrac{1}{3}ax^3 + \tfrac{1}{4}x^4$ \cite{LD15}.
The first example is $ \beta = 1$, $P= 0.59$, $a= -2$, and the second one $ \beta = 1$, $P = -0.5$, $a= 1.89$. In fact, the signature of the $\vec{G}$ matrices is identical to an even potential at $P=0$. Note that, in contrast to the even potential,
also the value of $\beta$ is fixed.

The cross-coupling leading to Phase 3 is discussed in \cite{StSp14,PSSS15}. An explicit example has been discovered only during the Les Houches summer school \cite{Me15} and reads
\begin{equation}\label{8.5}
V(x) = \tfrac{1}{2}x^2 + \cos(\pi (x-\tfrac{1}{3})) + \tfrac{1}{8}x^4\,.
\end{equation}
Then $G^1_{11} =10^{-8}$, $G^1_{-1-1} =0.164$, and $G^1_{00} =0.272$ at $P = 2.214$ and
$\beta = 1$.

\section{Low temperature DNLS}
\label{sec13}
We return to the DNLS introduced in Sect. \ref{sec8}. At high temperatures the conserved fields, density and energy, spread diffusively. However at low temperatures the field of phase differences is almost conserved, which allows for propagating modes.
According to the equilibrium measure,  for $\beta \gg 1$ and $\langle \rho_j\rangle =1$, the phase is similar to a discrete time random walk on $S^1$ with single step variance of order $\beta^{-1}$,
until the walk realizes the finite geometry and the variance crosses over to exponential decay. 
In one dimension there is no static phase transition, but in three and more dimensions a continuous symmetry can be broken. Then the field corresponding to such a broken symmetry has to be added to the list of conserved fields \cite{Fo75}. In a similar spirit, for DNLS at low temperatures the phase difference has to be included in the list of conserved fields.  

We adjust the chemical potential $\mu$ such that the average density $\bar{\rho}$ is fixed, $\bar{\rho}>0$. As one lowers the temperature, according to  the equilibrium measure, the $\rho_j$'s deviate from $\bar{\rho}$ by order $1/\sqrt{\beta}$ and also 
the phase difference $\vert\varphi_{j+1} - \varphi_{j}\rvert = \mathcal{O}(1/\sqrt{\beta}\,)$, while the phase itself is uniformly distributed over $[0,2\pi]$. We introduce $\tilde{r}_j = \Theta(\varphi_{j+1} -\varphi_j)$, where $\Theta$ is $2\pi$-periodic
and $\Theta(x) = x$ for $\lvert x \rvert \leq \pi$.  Since $\Theta$ has a jump discontinuity, $\tilde{r}_j $ is not conserved. In a more pictorial language, the event that $\lvert\varphi_{j+1}(t) -\varphi_j(t)\rvert =\pi$ is called an umklapp for phase difference $\tilde{r}_j$ or an umklapp process to emphasize its dynamical character. At low temperatures a jump of size $\pi$ has a small probability of order $e^{-\beta\Delta V}$ with $\Delta V $ the height of a suitable potential barrier still to be determined. Hence $\tilde{r}_j$ is locally conserved up to umklapp processes occurring with a very small frequency only, see \cite{DaDh14} for a numerical validation.

How to incorporate an almost conserved field into fluctuation hydrodynamics is not so obvious. Therefore we first 
construct an effective low-temperature  hamiltonian $H_\mathrm{lt}$, in  such a way that it has the same equilibrium measure and strictly conserves $\tilde{r}_j$. More concretely, we first parametrize the angles $\varphi_0, \dots, \varphi_{N-1}$ through $r_j = \varphi_{j+1} -\varphi_j$ with $r_j \in [-\pi,\pi]$. To distinguish, we denote the angles in this particular parametrization by $\phi_j$. $(\phi_j,\rho_j)$ is a pair of canonically conjugate variables. Umklapp is defined by $|r_j(t)| = \pi$. 
The dynamics governed by the DNLS Hamiltonian corresponds to periodic boundary conditions at $r_j = \pm \pi$. For  an approximate  low temperature description we impose instead specular reflection, \textit{i.e.}, if $r_j = \pm\pi$, then $\rho_j$, $\rho_{j+1}$ are scattered to $\rho_j' = \rho_{j+1}$, $\rho_{j+1}' = \rho_j$. By fiat, the approximate low temperature dynamics stricly conserves the phase difference $r_j$ and is  identical to the DNLS dynamics
between two umklapp events. The corresponding Hamiltonian then reads
\begin{equation}\label{9.1}
H_{\mathrm{lt}} = \sum_{j=0}^{N-1} \big( \sqrt{\rho_{j+1}\,\rho_j} \, U(\phi_{j+1} - \phi_j) + V(\rho_j) \big)\,,
\end{equation}
where
\begin{equation}\label{9.2}
U(x) = - \cos(x) \ \ \text{for}\ \ \lvert x \rvert \leq \pi \,,\qquad U(x) = \infty \ \ \text{for}\ \ \lvert x \rvert > \pi\,,
\end{equation}
and 
\begin{equation}\label{9.3}
V(x) =  x + \tfrac{1}{2} g\, x^2 \ \ \text{for}\ \ x \geq 0\,,\qquad V(x) = \infty \ \ \text{for}\ \ x < 0\,.
\end{equation}
As required before, the Boltzmann weights 
are not modified, $\exp[-\beta H] = \exp[-\beta H_{\mathrm{lt}}]$. For some computations it will be convenient to replace the hard collision potentials $U$, $V$ by a smooth variant, for which the infinite step is replaced by a rapidly diverging smooth potential. The dynamics is governed by
\begin{equation}\label{9.4}
\frac{d}{d t} \phi_j = - \partial_{\rho_j} H_{\mathrm{lt}}\,,\qquad\frac{d}{d t} \rho_j = \partial_{\phi_j} H_{\mathrm{lt}}\,,
\end{equation}
including the specular reflection of $\rho_j$ at $\rho_j= 0$ and of $r_j$ at $r_j = \pm \pi$. 
The solution trajectory agrees piecewise with the trajectory from the DNLS dynamics.
The break points are the umklapp events, which are extremely rare at low temperatures.

There are two potential barriers, $\Delta U$ and $\Delta V$. The minimum of $V(x) -x - \mu x$, $\mu >0$, is at $\bar{\rho} = \mu/g$, hence $\Delta V = \tfrac{1}{2}g\bar{\rho}^2$. The minimum of $U$ is at $\phi_{j+1} - \phi_j = 0$ and, setting $\rho_j = \bar{\rho}$, one arrives at $\Delta U = 2 \bar{\rho}$. Thus the low temperature regime is characterized by
\begin{equation}\label{9.5}
\tfrac{1}{2} \beta g\bar{\rho}^2 \gtrsim 1\,,\qquad 2 \beta \bar{\rho} \gtrsim 1\,.
\end{equation}
In this parameter regime we expect the equilibrium time correlations based on $H_{\mathrm{lt}}$ to well approximate the time correlations of the exact DNLS.

The conserved fields are now $\rho_j$, $r_j$, and the energy
\begin{equation}\label{9.6}
e_j = \sqrt{\rho_{j+1}\,\rho_j}\,U(r_j) + V(\rho_j) \,.
\end{equation}
This local energy differs from the one introduced in \eqref{4.4} by the term $\tfrac{1}{2}(\rho_{j+1} - \rho_j)$. In the expressions below such a difference term drops out and in the final result we could use either one. The local conservation laws and their currents read, for the density
\begin{equation}\label{9.7}
\frac{d}{d t} \rho_j + \mathcal{J}_{\rho,j+1} - \mathcal{J}_{\rho,j} = 0
\end{equation}
with local density current
\begin{equation}\label{9.8}
\mathcal{J}_{\rho,j} = \sqrt{\rho_{j-1}\,\rho_{j}} \, U'(r_{j-1})\,,
\end{equation}
for the phase difference
\begin{equation}\label{9.9}
\frac{d}{d t} r_j + \mathcal{J}_{r,j+1} - \mathcal{J}_{r,j} = 0
\end{equation}
with local phase difference current
\begin{equation}\label{9.10}
\mathcal{J}_{r,j} = \tfrac{1}{2} \sqrt{\rho_{j+1}/\rho_j}\,U(r_j) + \tfrac{1}{2} \sqrt{\rho_{j-1}/\rho_j}\,U(r_{j-1}) + V'(\rho_j)\,,
\end{equation}
and for the energy
\begin{equation}\label{9.10a}
\frac{d}{d t} e_j + \mathcal{J}_{e,j+1} - \mathcal{J}_{e,j} = 0
\end{equation}
with local energy current
\begin{equation}\label{9.11}
\mathcal{J}_{e,j} = \tfrac{1}{2} \sqrt{\rho_{j-1}\,\rho_{j+1}}\,\big( U(r_{j-1})U'(r_j) + U'(r_{j-1}) U(r_j) \big) + \sqrt{\rho_{j-1}\,\rho_j}\,U'(r_{j-1}) V'(\rho_j) \, .
\end{equation}
To shorten notation, we set $\vec{g}_j = (\rho_j,r_j,e_j)$ and $\vec{\mathcal{J}}_j = \big( \mathcal{J}_{\rho,j}, \mathcal{J}_{r,j}, \mathcal{J}_{e,j} \big)$.

We follow the blue-print provided by the anharmonic chains.  The Euler currents are determined as the thermal average of the microscopic currents.
The canonical state has as parameters $\beta$, $\mu$ and, in addition, a second  chemical potential, $\nu$,
which sets the average phase difference $\langle r_j\rangle$. Physically the phase difference cannot be controlled, hence $\nu = 0$.
However for the second order expansion one has to first take derivatives and then set $\nu = 0$. At the first sight it looks difficult to draw any conclusions on the coupling matrices $\vec{G}$. There is  however a surprising identity which comes for rescue,
\begin{equation}\label{9.12}
\big\langle \vec{\mathcal{J}_j} \big\rangle_{\beta,\mu,\nu} = \langle ( \mathcal{J}_{\rho,j}, \mathcal{J}_{r,j}, \mathcal{J}_{e,j} ) 
\rangle_{\beta,\mu,\nu}  = (\nu, \mu, \mu\,\nu) = \vec{\mathsf{j}}\,,
\end{equation}
which should be compared with \eqref{5.1}. A tricky computation is still ahead.  But at the end one finds that 
$G^0_{00} = 0$ and $G^1_{11} > 0$, while $G^1$ equals $-G^{-1}$ transposed relative to the anti-diagonal.  

Of course, the equations of nonlinear fluctuating hydrodynamics have the same structure as explained before.
Thus we conclude that the dynamical phase of low temperature DNLS is identical to the  generic Phase 1 of an anharmonic chain.

\section{Mode-coupling theory}
\label{sec14}
\textbf{Decoupling hypothesis}. For the linear Langevin equations the normal modes decouple for long times. As first argued by van Beijeren
\cite{vB12} 
such decoupling persists when adding the quadratic nonlinearities. For the precise phrasing, we have to be somewhat careful. We consider a fixed component, $\alpha$, in normal mode representation. It travels with velocity $c_\alpha$, which is assumed to be distinct from all other mode velocities. If $G^\alpha_{\alpha\alpha} \neq 0$, then for the purpose of computing correlations of mode $\alpha$ at large scales, one can use the scalar conservation law
\begin{equation}\label{B.1a}
\partial_t \phi_\alpha + \partial_x \big(c_\alpha \phi_\alpha + G^{\alpha}_{\alpha\alpha} \phi_\alpha^2-D^\sharp_{\alpha\alpha}\partial_x\phi_\alpha + B^\sharp_{\alpha\alpha}\xi_\alpha\big)=0\,,
\end{equation}
which coincides with the stochastic Burgers equation \eqref{1.20}. If decoupling holds, one has the exact asymptotics as stated in \eqref{1.24} with $\lambda = 2G^\alpha_{\alpha\alpha}$. 

As discussed in Sect. \ref{sec12}, for a generic anharmonic chain $G^1_{11} = -G^{-1}_{-1-1}\neq 0$. Hence,
if  $G^1_{11} \neq 0$, the decoupling hypothesis asserts that the \textit{exact} scaling form of the sound peak is
\begin{equation}\label{B.2a}
S^{\sharp}_{\sigma\sigma}(x,t) \simeq 2^{-1}(\Gamma_\mathrm{s} t)^{-2/3} f_{\mathrm{KPZ}}
\big(2^{-1}(\Gamma_\mathrm{s} t)^{-2/3}(x - \sigma c t)\big)\,,\quad \Gamma_\mathrm{s} = |G^\sigma_{\sigma\sigma}|\,,
 \end{equation}
$\sigma = \pm 1$. On the other hand $G_{00}^0=0$ always.
To find out about the scaling behavior of the heat peak other methods have to be developed. 
\medskip\\
\textbf{One-loop, diagonal, and small overlap approximations}. 
 The Langevin equation~(\ref{7.5}) is slightly formal. To have a well-defined evolution, we discretize space by a lattice of $N$ sites. The field $\vec{\phi}(x,t)$ then becomes $\vec{\phi}_j(t)$ with components $\phi_{j,\alpha}(t)$, $j=1,\ldots,N$, $\alpha=0,\pm1$. The spatial  finite difference operator is denoted by $\partial_j$, $\partial_j f_j=f_{j+1}-f_{j}$, with transpose $\partial_j^\mathrm{T} f_j=f_{j-1}-f_{j}$. Then the discretized equations of fluctuating hydrodynamics read
\begin{equation}\label{B.1}
\partial_t \phi_{j,\alpha}+\partial_j\big(c_\alpha \phi_{j,\alpha} +\mathcal{N}_{j,\alpha} + \partial_j^\mathrm{T} D^\sharp\phi_{j,\alpha} + B^\sharp\xi_{j,\alpha}\big)=0
\end{equation}
with $\vec{\phi}_{j}=\vec{\phi}_{N+j}$, $\vec{\xi}_0=\vec{\xi}_N$, where $\xi_{j,\alpha}$ are independent Gaussian white noises with covariance
\begin{equation}\label{B.2}
\langle \xi_{j,\alpha}(t) \xi_{j',\alpha'} (t')\rangle =\delta_{jj'} \delta_{\alpha\alpha'} \delta(t-t')\,.
\end{equation}
The diffusion matrix $D^\sharp$ and noise strength $B^\sharp$ act on components, while the difference operator $\partial_j$ acts on the lattice site index $j$.

$\mathcal{N}_{j,\alpha}$ is quadratic in $\phi$. But let us first consider the case $\mathcal{N}_{j,\alpha} =0$. Then $\phi_{j,\alpha}(t)$ is a Gaussian process. The noise strength has been chosen such that one invariant measure is the Gaussian
\begin{equation}\label{B.3}
\prod^N_{j=1} \prod_{\alpha=0,\pm1} \exp[-\tfrac{1}{2}\phi^2_{j,\alpha}] (2\pi)^{-1/2} d \phi_{j,\alpha}= \rho_\mathrm{G} (\phi) \prod^N_{j=1} \prod_{\alpha=0,\pm1} d \phi_{j,\alpha}\,.
\end{equation}
Because of the conservation laws, the hyperplanes
\begin{equation}\label{B.4}
\sum^N_{j=1} \phi_{j,\alpha}=N\rho_\alpha\,,
\end{equation}
are invariant and on each hyperplane there is a Gaussian process with a unique invariant measure given by
 (\ref{B.3}) conditioned on that hyperplane.
For large $N$ it would become independent Gaussians with mean $\rho_\alpha$, our interest being the case of zero
mean,  $\rho_\alpha=0$. 

The generator of the diffusion process~(\ref{B.1}) with $\mathcal{N}_{j,\alpha}=0$ is given by
\begin{equation}\label{B.5}
L_0=\sum^N_{j=1} \Big(-\sum_{\alpha=0,\pm 1} \partial_j\big(c_\alpha \phi_{j,\alpha}+ \partial_j^\mathrm{T} D^\sharp\phi_{j,\alpha}\big) \partial_{\phi_{j,\alpha}}+\sum_{\alpha,\alpha'=0,\pm1}  (B^\sharp B^{\sharp\mathrm{T}})_{\alpha\alpha'}  \partial_j \partial_{\phi_{j,\alpha}} \partial_j\partial_{\phi_{j,\alpha'}}\Big)\,.
\end{equation}
The invariance of $\rho_\mathrm{G}(\phi)$ can be checked through
\begin{equation}\label{B.7}
L^\ast_0 \rho_\mathrm{G}(\phi)=0\,,
\end{equation}
where $^\ast$ is the adjoint with respect to the flat volume measure. Furthermore linear functions evolve to linear functions according to
\begin{equation}\label{B.6}
\mathrm{e}^{L_0 t}\phi_{j,\alpha}= \sum^N_{j'=1} \sum_{\alpha'=0,\pm1} (\mathrm{e}^{\mathcal{A}t})_{j\alpha,j'\alpha'} \phi_{j',\alpha'}\,,
\end{equation}
where the matrix $\mathcal{A}=- \partial_j\otimes \mathrm{diag} (-c,0,c) - \partial_j \partial_j^\mathrm{T} \otimes D^\sharp$, the first factor acting on $j$ and the second on $\alpha$.

We now add the nonlinearity $\mathcal{N}_{j,\alpha}$. In general, this will modify the time-stationary measure and we have little control on how. Therefore we propose to choose $\mathcal{N}_{j,\alpha}$ such that the corresponding vector field
$\partial_j\mathcal{N}_{j,\alpha}$ is divergence free \cite{SaSp09}. If $\mathcal{N}_{j,\alpha}$ depends only on the field at sites $j$ and $j+1$, then the unique solution reads
\begin{equation}\label{B.8}
\mathcal{N}_{j,\alpha}=\tfrac{1}{3}\sum_{\gamma,\gamma'=0,\pm1} G^\alpha_{\gamma\gamma'}\big(\phi_{j,\gamma}\phi_{j,\gamma'}+\phi_{j,\gamma}\phi_{j+1,\gamma'}+\phi_{j+1,\gamma}\phi_{j+1,\gamma'}\big)\,.
\end{equation}
For $\rho_\mathrm{G}$ to remain invariant under 
the deterministic flow generated by the vector field $-\partial_j\mathcal{N}$ requires
\begin{equation}\label{B.9}
L_1^\ast\rho_\mathrm{G} = 0\,,\qquad L_1= - \sum^N_{j=1} \sum_{\alpha=0,\pm 1} \partial_j \mathcal{N}_{j,\alpha} \partial_{\phi_{j,\alpha}}\,,
\end{equation}
which implies
\begin{equation}\label{B.10}
\sum^N_{j=1} \sum_{\alpha=0,\pm1} \phi_{j,\alpha} \partial_j \mathcal{N}_{j,\alpha}=0
\end{equation}
and thus the cyclicity constraints
\begin{equation}\label{B.9a}
G^\alpha_{\beta\gamma}=G^\beta_{\alpha\gamma} \,\big(=G^\alpha_{\gamma\beta}\big)
\end{equation}
for all $\alpha,\beta,\gamma=1,2,3$, where in brackets we added the symmetry which holds by definition. Denoting the generator of the Langevin equation~(\ref{B.1}) by
\begin{equation}\label{B.10a}
  L=L_0+L_1\,,
\end{equation}
one concludes $L^\ast \rho_\mathrm{G}=0$, \textit{i.e.} the time-invariance of $\rho_\mathrm{G}$. The cyclicity condition also appears for coupled KPZ equations \cite{Fu15}.

Unfortunately, the cyclicity condition fails for anharmonic chains. On the other hand, to derive the mode-coupling equations the explicit form of the stationary measure seems to be required. To avoid a deadlock we argue by universality. First, the relevant couplings $G^\alpha_{\alpha'\alpha'}$ are computed from the particular anharmonic chain.
We then fix all other couplings by cyclicity. Of course the true irrelevant couplings are different. But this will not show in the long time behavior. 

We return to the Langevin equation (\ref{B.1}) and consider the mean zero, stationary $\phi_{j,\alpha}(t)$ process
with $\rho_\mathrm{G}$ as $t=0$ measure.  The stationary covariance reads
\begin{equation}\label{B.14}
S^\sharp_{\alpha\alpha'}(j,t)=\langle \phi_{j,\alpha}(t)\phi_{0,\alpha'}(0)\rangle =\langle \phi_{0,\alpha'}\mathrm{e}^{Lt}\phi_{j,\alpha}\rangle_{\mathrm{eq}}\,,\quad t \geq 0\,.
\end{equation}
On the left, $\langle\cdot\rangle$ denotes the average with respect to the stationary $\phi_{j,\alpha}(t)$ process and on the right 
$\langle\cdot\rangle_{\mathrm{eq}}$ refers to the average with respect to $\rho_\mathrm{G}$.
By construction
\begin{equation}\label{B.15}
S^\sharp_{\alpha\alpha'}(j,0)=\delta_{\alpha\alpha'} \delta_{j0}\,.
\end{equation}
The time derivative reads
\begin{equation}\label{B.16}
\frac{d}{dt} S^\sharp_{\alpha\alpha'}(j,t)=\langle \phi_{0,\alpha'}(\mathrm{e}^{Lt}L_0 \phi_{j,\alpha})\rangle_{\mathrm{eq}}+ \langle \phi_{0,\alpha'}(\mathrm{e}^{Lt}L_1 \phi_{j,\alpha})\rangle_{\mathrm{eq}}\,.
\end{equation}
We insert
\begin{equation}\label{B.17}
\mathrm{e}^{Lt}=\mathrm{e}^{L_0 t}+\int^t_0 ds\, \mathrm{e}^{L_0(t-s)} L_1 \mathrm{e}^{Ls}
\end{equation}
in the second summand of~(\ref{B.16}). The term containing only $\mathrm{e}^{L_0 t}$ is cubic in the time zero fields and hence its average vanishes.  Therefore one arrives at 
\begin{equation}\label{B.18}
\frac{d}{dt} S^\sharp_{\alpha\alpha'}(j,t)=  \mathcal{A} S_{\alpha\alpha'}(j,t) + \int^t_0 ds \langle \phi_{0,\alpha'} \mathrm{e}^{L_0(t-s)}L_1(\mathrm{e}^{Ls} L_1 \phi_{j,\alpha}) \rangle_\mathrm{eq}\,.
\end{equation}
For the adjoint of $\mathrm{e}^{L_0(t-s)}$ we use~(\ref{B.6}) and for the adjoint of $L_1$ we use
\begin{equation}\label{B.18a}
\langle \phi_{j,\alpha}  L_1  F(\phi)\rangle_\mathrm{eq} =  - \langle (L_1 \phi_{j,\alpha})   F(\phi)\rangle_\mathrm{eq} \,,
\end{equation}
which both rely on $\langle \cdot \rangle_\mathrm{eq}$ being the average with respect to $\rho_\mathrm{G}$. Furthermore
\begin{equation}\label{B.19}
L_1 \phi_{j,\alpha} = - \partial_j \mathcal{N}_{j,\alpha}\,.
\end{equation}
 Inserting in~(\ref{B.18}) one arrives at the identity
\begin{equation}\label{B.18b}
 \frac{d}{dt} S^\sharp_{\alpha\alpha'}(j,t)= \mathcal{A}S^\sharp_{\alpha\alpha'}(j,t)
  - \int^t_0 ds  \langle
(\mathrm{e}^{\mathcal{A}^\mathrm{T}(t-s)}\partial_j \mathcal{N}_{0,\alpha'})(\mathrm{e}^{Ls} \partial_j \mathcal{N}_{j,\alpha}) \rangle_\mathrm{eq}\,.
\end{equation}

To obtain a closed equation for $S^\sharp$ we note that the average 
\begin{equation}\label{B.19a}
\langle \partial_{j'}  \mathcal{N}_{j',\alpha'} \mathrm{e}^{L_s} \partial_j 
\mathcal{N}_{j,\alpha}\rangle_\mathrm{eq} = \langle \partial_{j } \mathcal{N}_{j,\alpha}(s)\partial_{j'} 
\mathcal{N}_{j',\alpha'}(0)\rangle
\end{equation}
is
a four-point correlation. We invoke the Gaussian factorization as
\begin{equation}\label{B.20a}
\langle\phi(s)\phi(s)\phi(0)\phi(0)\rangle\cong \langle\phi(s)\phi(s)\rangle\langle\phi(0)\phi(0)\rangle+2\langle\phi(s)\phi(0)\rangle\langle\phi(s)\phi(0)\rangle\,.
\end{equation}
 The first summand vanishes because of the difference operator $\partial_j$. Secondly we replace the bare propagator $\mathrm{e}^{\mathcal{A}(t-s)}$ by the interacting propagator $S^\sharp(t-s)$, which corresponds to a partial resummation of the perturbation series in $\vec{G}$. Finally we take a limit of zero lattice spacing. This step could be avoided, and is done so in our numerical scheme for the mode-coupling equations. We could also maintain the ring geometry which,
  for example,  would allow 
 to investigate  collisions between the moving peaks. Universality is only expected for large $j,t$, hence in the limit of zero lattice spacing. The continuum limit of $S^\sharp(j,t)$ is denoted by $S^\sharp(x,t)$, $x\in\mathbb{R}$. With these steps we arrive at the mode-coupling equation
 \begin{eqnarray}\label{B.20}
&&\hspace{-53pt}  \partial_t S^\sharp_{\alpha\beta}(x,t)= \sum_{\alpha'=0,\pm1} \Big(\big(-c_\alpha\delta_{\alpha\alpha'}\partial_x +D_{\alpha\alpha'}\partial^2_x\big) S^\sharp_{\alpha'\beta}(x,t)\nonumber\\
&&\hspace{45pt} + \int^t_0 ds \int_{\mathbb{R}} dy    M_{\alpha\alpha'}(y,s) \partial^2_xS^\sharp_{\alpha'\beta}(x-y,t-s)\Big)
\end{eqnarray}
with the memory kernel
\begin{equation}\label{B.21}
M_{\alpha\alpha'}(x,t)= 2\sum_{\beta',\beta'',\gamma',\gamma''=0,\pm1} G^\alpha_{\beta'\gamma'} G^{\alpha'}_{\beta''\gamma''} S^\sharp_{\beta'\beta''}(x,t) S^\sharp_{\gamma'\gamma''}(x,t)\,.
\end{equation}

In numerical simulations of both, the mechanical model of anharmonic chains and the mode-coupling equations,  it is consistently observed that $S^{\sharp}_{\alpha\alpha'}(j,t)$ becomes  approximately  diagonal fairly rapidly. To analyse the long time asymptotics on the basis of \eqref{B.20} we therefore rely on the diagonal approximation
\begin{equation}\label{40}
  S^{\sharp}_{\alpha\alpha'}(x,t)\simeq \delta_{\alpha\alpha'} f_\alpha(x,t)\,.
\end{equation}
Then $f_\alpha(x,0)=\delta(x)$ and the $f_\alpha$'s satisfy
\begin{equation}\label{41}
\partial_t f_\alpha(x,t)= (-c_\alpha \partial_x+D^\sharp_{\alpha\alpha} \partial^2_x) f_\alpha (x,t) + \int^t_0 ds \int_{\mathbb{R}} dy
 \partial^2_x  f_\alpha(x-y,t-s) M_{\alpha\alpha}(y,s)\,,
\end{equation}
$\alpha=-1,0,1$,  with memory kernel
\begin{equation}\label{41a}
M_{\alpha\alpha}(x,t)=2 \sum_{\gamma,\gamma'=0,\pm 1} (G^\alpha_{\gamma\gamma'})^2 f_\gamma(x,t) f_{\gamma'}(x,t)\,.
\end{equation}

The solution to \eqref{41} has two sound peaks centered at 
$\pm ct$ and the heat peak  sitting at $0$. All three peaks  have a width much less than $ct$. But then,
in case $\gamma\neq \gamma'$, the product  $f_\gamma(x,t) f_{\gamma'}(x,t) \simeq 0$ for large $t$. Hence for the memory kernel \eqref{41a} we invoke a small overlap approximation as
\begin{equation}\label{43}
M_{\alpha\alpha}(x,t)\simeq M^{\mathrm{dg}}_{\alpha}(x,t)=2\sum_{\gamma=0,\pm 1}(G^\alpha_{\gamma\gamma})^2 f_\gamma(x,t)^2\,,
\end{equation}
which is to be inserted in Eq. \eqref{41}. 

The decoupling hypothesis can also be applied to Eqs. \eqref{41} and \eqref{43}.  In fact, in the summer 2012 C. Mendl implemented numerical solutions of the full mode-coupling equations \eqref{B.20} on fairly small lattices. The decoupling is seen very convincingly. One then arrives at the mode-coupling equation for a single component, written down already in 1985 \cite{vBKS85}. In the long time limit the spreading is proportional to  $t^{2/3}$ and the corresponding scaling function deviates only by approximately 5\% from the exact scaling function \eqref{B.2a}.

We now have a tool available, by which also the heat peak can be handled, at least approximately. Setting $\alpha = 0$ in \eqref{41} with approximation \eqref{43},
$f_0$ couples to $f_{-1}f_{-1}$ and $f_{1}f_{1}$. But these are presumably close to $f_\mathrm{KPZ} $ of Eq. \eqref{B.2a}. In fact, the scaling exponent matters while the precise shape  modifies only prefactors. Inserting 
$f_\mathrm{KPZ} $ in  Eq. \eqref{41} and solving the resulting linear equation for $f_0$, one obtains  that for long times the Fourier transform of $f_0$ is given by
\begin{equation}\label{50}
\hat{f}_0(k,t) = \mathrm{e}^{-|k|^{5/3} |\Gamma_\mathrm{h} t|}\,.
\end{equation}
There is a somewhat lengthy formula for $\Gamma_\mathrm{h} $, see Eq. (4.12) of \cite{Sp14}. The right hand side is the Fourier transform of the L\'{e}vy probability distribution with exponents $\alpha = \tfrac{5}{3}, \beta =0$.
The tail of the L\'{e}vy $\tfrac{5}{3}$ distribution decays $|x|^{-8/3}$, which has no second moment. But in fact the distribution is cut off at the sound peaks. There are no correlations propagating beyond the sound cone. 

Besides the generic dynamical Phase 1, mode-coupling covers also the other dynamical phases. For Phase 2, the sound peaks are diffusive with scaling function
\begin{equation}\label{44}
f_\sigma(x,t)=\frac{1}{\sqrt{4\pi D_\mathrm{s} t}}\mathrm{e}^{-(x-\sigma ct)^2/4 D_\mathrm{s} t}\,.
\end{equation}
$D_\mathrm{s}$ is a transport coefficient. It can defined through a Green-Kubo formula, which also means that no reasonably explicit answer can be expected. 
The feed back of the sound peak to the central peak follows by the same steps as before, with the result   
\begin{equation}\label{45}
\hat{f}_0(k,t)=\mathrm{e}^{-|k|^{3/2} \Gamma{_\mathrm{h}}|t|}\,,
\end{equation}
where
\begin{equation}\label{8.6}
\Gamma_\mathrm{h}= (D_\mathrm{s})^{-1/2} (G^0_{\sigma\sigma})^2 (4\pi)^2(2\pi c)^{-1/2} \int^\infty_0 dt \,t^{-1/2} \cos (t)
(2\sqrt{\pi})^{-1}\,.
\end{equation}
Since $\tfrac{3}{2} < \tfrac{5}{3}$, the density $f_0(x,t)$ turns out to be broader than the L\'{e}vy $\tfrac{5}{3}$ from the Phase 1.  

For Phase 3, one proceeds as before, but leaves the scaling exponent yet undetermined. 
Since modes are cross-coupled,  one repeats the argument with another pair of modes. The only solution turns out to be L\'{e}vy with the golden mean $\alpha = \tfrac{1}{2} ( 1 + \sqrt{5})$. (In the theory of stable laws
the two relevant parameters are usually denoted by $\alpha,\beta$, see \cite{UZ99}).
However one also picks up phase factors.
This then yields  the asymmetry parameters $\beta= -1,0,1$, respectively for each peak, consistent with the physical principle of no correlations beyond the sound cone. 

\section{Molecular dynamics and other missed topics}
\label{sec15}
Anomalous transport in one-dimensional chains is a fascinating topic with many contributions, including extensive molecular dynamics simulations, which in absence of accurate experiments is the only mean to check theoretical predictions. As with other items, I cannot provide any details here. Instead I refer to the 2003 article by S. Lepri,
R. Livi, and
A. Politi \cite{LeLi03} and the 2008 article by A. Dhar \cite{Dh08}. These reviews extensively cover  the popular scheme of coupling the chain at its two ends to thermal reservoirs of different temperatures. While, in principle, nonlinear fluctuating hydrodynamics should cover also such boundary conditions, no progress has been achieved
yet in this direction. 
Also I had to omit a discussion of the current-current correlations in equilibrium  \cite{MeSp14}. There are recent attempts to
go beyond covariances and,
 in spirit of the progress on the KPZ equation, to uncover the full probability density function of the time-integrated currents. Over the past few years several molecular dynamics simulations  have been performed
 with the goal to check the validity of nonlinear fluctuating hydrodynamics. A fairly complete discussion can be found 
in  my contribution to a forthcoming volume of Springer Lecture Notes in Physics, edited by S. Lepri, on thermal transport in low dimensions \cite{Sp16}. \\\\
\textbf{Acknowledgements}: In my explorations of the KPZ landscape I had the good luck of benefitting from outstanding collaborators, starting with Henk van Beijeren and Joachim Krug in the early days and more recently with Michael Pr\"{a}hofer, Patrik Ferrari, Tomohiro Sasamoto, Sylvain Prolhac, and Thomas Weiss.
For nonlinear fluctuating hydrodynamics I highly appreciate the cooperation with Christian Mendl.
I enjoyed tremendously the stimulating atmosphere at Les Houches.

\end{document}